\pdfoutput=1

\documentclass[11pt]{article}

\usepackage[preprint]{acl}

\usepackage{times}
\usepackage{latexsym}

\usepackage[T1]{fontenc}

\usepackage[utf8]{inputenc}

\usepackage{microtype}

\usepackage{inconsolata}
\usepackage{xcolor}         
\usepackage{graphicx} 
\usepackage{subfigure}
\usepackage[most,breakable]{tcolorbox}
\usepackage[dvipsnames*,svgnames]{xcolor}
\usepackage{wrapfig}
\usepackage[normalem]{ulem}
\usepackage{tcolorbox}
\usepackage{caption}
\usepackage{enumitem}
\usepackage{color}
\usepackage{hyperref}
\usepackage{authblk}
\useunder{\uline}{\ul}{}
\definecolor{gold}{RGB}{255,215,0}
\definecolor{silver}{RGB}{192,192,192}
\definecolor{bronze}{RGB}{205,127,50}

\definecolor{boxcolor1}{RGB}{235,240,255} 
\definecolor{boxcolor2}{RGB}{255,240,230} 
\definecolor{boxcolor3}{RGB}{230,255,230} 
\definecolor{boxcolor4}{RGB}{255,230,255} 
\definecolor{boxcolor5}{RGB}{255,255,230} 
\definecolor{boxcolor6}{RGB}{230,255,255} 
\definecolor{grey}{RGB}{240,240,240} 

\newtcolorbox{questionbox}[1][]{
  colback=#1,
  colframe=black,
  boxrule=1pt,
  arc=0pt,
  enhanced,
  breakable
}
\newtcolorbox{examplebox}[2][]{
  colback=#1,
  colframe=black,
  boxrule=1pt,
  enhanced,
  breakable,
  title=#2,            
  coltitle=black,      
  colbacktitle=gray!20,
  fonttitle=\bfseries  
}
\title{AstroMMBench: A Benchmark for Evaluating Multimodal Large Language Models Capabilities in Astronomy}

\author{
  Jinghang Shi$^{1,2}$  
  Xiaoyu Tang$^{3}$  
  Yang Huang$^{1,2}$  
  Yuyang Li$^{1,2}$  
  Xiao Kong$^{1}$ 
  Yanxia Zhang$^{1}$ 
  Caizhan Yue$^{4}$
  \\
  $^{1}$University of Chinese Academy of Sciences \\
  $^{2}$National Astronomical Observatories, Chinese Academy of Sciences\\
  $^{3}$Zhejiang Laboratory\\
  $^{4}$Tianjin University\\
}

\begin{document}
\maketitle
\begin{abstract}
 Astronomical image interpretation presents a significant challenge for applying multimodal large language models (MLLMs) to specialized scientific tasks. Existing benchmarks focus on general multimodal capabilities but fail to capture the complexity of astronomical data. To bridge this gap, we introduce \textbf{AstroMMBench}, the first comprehensive benchmark designed to evaluate MLLMs in astronomical image understanding. AstroMMBench comprises 621 multiple-choice questions across six astrophysical subfields, curated and reviewed by 15 domain experts for quality and relevance. We conducted an extensive evaluation of 25 diverse MLLMs, including 22 open-source and 3 closed-source models, using AstroMMBench. The results show that Ovis2-34B achieved the highest overall accuracy (70.5\%), demonstrating leading capabilities even compared to strong closed-source models. Performance showed variations across the six astrophysical subfields, proving particularly challenging in domains like cosmology and high-energy astrophysics, while models performed relatively better in others, such as instrumentation and solar astrophysics. These findings underscore the vital role of domain-specific benchmarks like AstroMMBench in critically evaluating MLLM performance and guiding their targeted development for scientific applications. AstroMMBench provides a foundational resource and a dynamic tool to catalyze advancements at the intersection of AI and astronomy.
\end{abstract}

\section{Introduction}

Astronomy is a field that relies heavily on observation. The analysis and interpretation of telescope-collected image data is a crucial method for astronomers to understand the universe. The increasing volume and complexity of astronomical data, driven by advanced telescopic technologies, pose increasing challenges for efficient and accurate data interpretation. Consequently, the quest for more advanced image analysis technologies has consistently been a significant direction in astronomical research.

Recently, as large language models (LLMs) \cite{devlin2018bert, Brown2020LanguageMA, Zeng2022GLM130BAO,  Bai2023QwenTR, Dubey2024Llama31, deepseekai2024deepseekv3} and large visual models (LVMs) \cite{Ramesh2021ZeroShotTG, Zhang2022DINODW, Kirillov2023SegmentA, Shen2023AligningAP, zhai2023sigmoid,fini2024AIMv2e,chen2024expanding} have been advancing, researchers have increasingly acknowledged the synergy effects that exist between these two types of models. This recognition has accelerated the formation and advancement of multimodal large language models (MLLMs) \cite{Achiam2023GPT4TR, wang2023cogvlm, yao2024minicpm, tong2024cambrian1, Abdin2024Phi3TR, liu2024llavanext, glm2024chatglm, bai2025qwen2, wu2024deepseekvl2, gemma_2025, kimiteam2025kimivl, zhu2025internvl3, dong2025scalable}. MLLMs combine the advanced natural language processing capabilities of LLMs with the visual comprehension strengths of LVMs, enabling them to possess both extensive world knowledge and advanced abilities in solving general visual tasks and complex reasoning \cite{Huang2024FromLL}. This combination of capabilities allows MLLMs to perform deeper and more detailed analysis of text and images, showing significant potential and value across various domains, such as healthcare  \cite{Guo2024PerformanceEO}, autonomous driving \cite{cui2023autodrive}, and art \cite{Ko2022LargescaleTG}.

It is foreseeable that MLLMs, with their powerful visual perception and understanding capabilities, will have enormous potential to assist astronomers in analyzing astronomical observation images. However, evaluating the performance of MLLMs in astronomical image understanding remains challenging. Although there are many multimodal benchmarks \cite{yue2024mmmu, chen2024mmstar, Wang2024CharXiv, Masry2022ChartQA, Li2023SEEDBench2, Huang2024AesBench, lu2024mathvista} available for evaluating the performance of MLLMs, they focus either on the models' comprehensive capabilities or on specific nonastronomical tasks. These benchmarks lack the domain specificity needed to assess a model's ability to handle tasks that require specialized knowledge of astrophysical processes. 

To address this gap, we introduce \textbf{AstroMMBench}, the first benchmark specifically designed to evaluate the performance of MLLMs in astronomy.  AstroMMBench includes 621 multiple-choice questions generated through an automated pipeline using images of papers on arxiv\footnote{https://arxiv.org/}, which have been rigorously vetted by 15 domain experts. These questions span six major subfields, from Galactic Astrophysics to Cosmology, providing a comprehensive framework for assessing capabilities in the field of MLLMs astronomy.

We evaluated 25 diverse MLLMs, comprising 22 publicly available open-source and 3 powerful closed-source models, using the VLMEvalKit framework and found significant performance differences across models and different subfields of astronomy. The results indicate that Ovis2-34B \cite{lu2024ovis} performs particularly well in various astrophysical tasks, achieving an overall score of 70.53\%. Notably, its performance surpassed that of leading closed-source models like ChatGPT-4o \cite{hurst2024gpt} (69.07\%) and Doubao-1.5-vision-pro (68.12\%), demonstrating the strong competitiveness of open-source solutions in professional field tasks. These findings underscore the importance of domain-specific benchmarks for advancing MLLMs in scientific research. We hope that AstroMMBench will become a key tool at the intersection of astronomy and artificial intelligence, promoting the development of models with better astronomical image understanding capabilities.

\section{Related Work}
\begin{figure}
    \centering
    \includegraphics[width=0.9\linewidth]{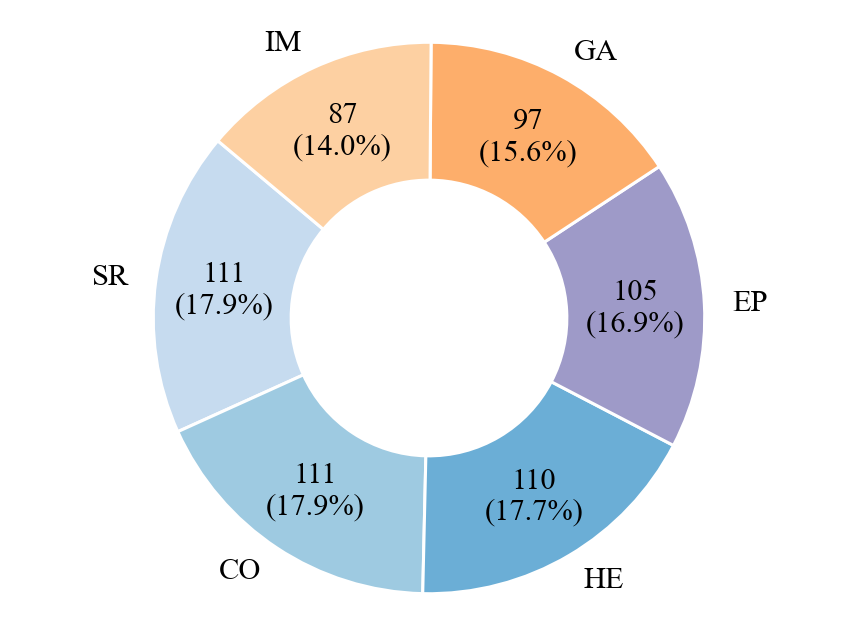}
    \caption{Distribution of questions across astronomy subfields in AstroMMBench.}
    \label{fig:categories-distribution}
\end{figure}
Evaluating the diverse capabilities of MLLMs necessitates comprehensive benchmarks. Existing general multimodal benchmarks \cite{yu2024mm, chen2024mmstar, yue2024mmmu, mmtbench, song2024milebench,li2024llava,Li2023SEEDBench2, fu2024mme}, primarily focus on everyday scenarios and common knowledge. They cover tasks such as image captioning, visual question answering (VQA), object perception, and complex reasoning across more than 20 skill dimensions. While these benchmarks are essential for measuring fundamental multimodal abilities and general world knowledge, they typically rely on common image types and scenarios, thus lacking the specialized content and nuanced understanding required for performance evaluation in fine-grained scientific domains.

To address the limitations of general-purpose benchmarks in terms of domain-specific knowledge coverage and task complexity, researchers have developed an increasing number of specialized evaluation suites. In the domain of mathematical and logical reasoning, benchmarks such as MathVista \cite{lu2024mathvista}, MathVerse \cite{zhang2024mathverse}, and We-Math \cite{qiao2024we} have been introduced to assess models' capabilities in understanding and solving visually presented mathematical problems. For chart and diagram understanding, datasets like ChartQA \cite{Masry2022ChartQA}, ChartX \cite{xia2024chartx}, and CharXiv \cite{Wang2024CharXiv} focus on evaluating model performance in chart recognition and complex reasoning tasks. Significant progress has also been made in multimodal evaluation for the medical domain, where benchmarks such as MedXpertQA \cite{zuo2025medxpertqa} and MediConfusion \cite{sepehri2024mediconfusion} systematically examine model performance in medical image diagnosis, pathology recognition, and other critical clinical tasks. The emergence of these domain-specific benchmarks has significantly advanced the evaluation of MLLMs in complex, specialized scenarios, offering a standardized framework for fine-grained capability analysis and promoting their applicability in high-stakes, expert-driven contexts.
\begin{figure*}
    \centering
    \includegraphics[width=0.8\linewidth]{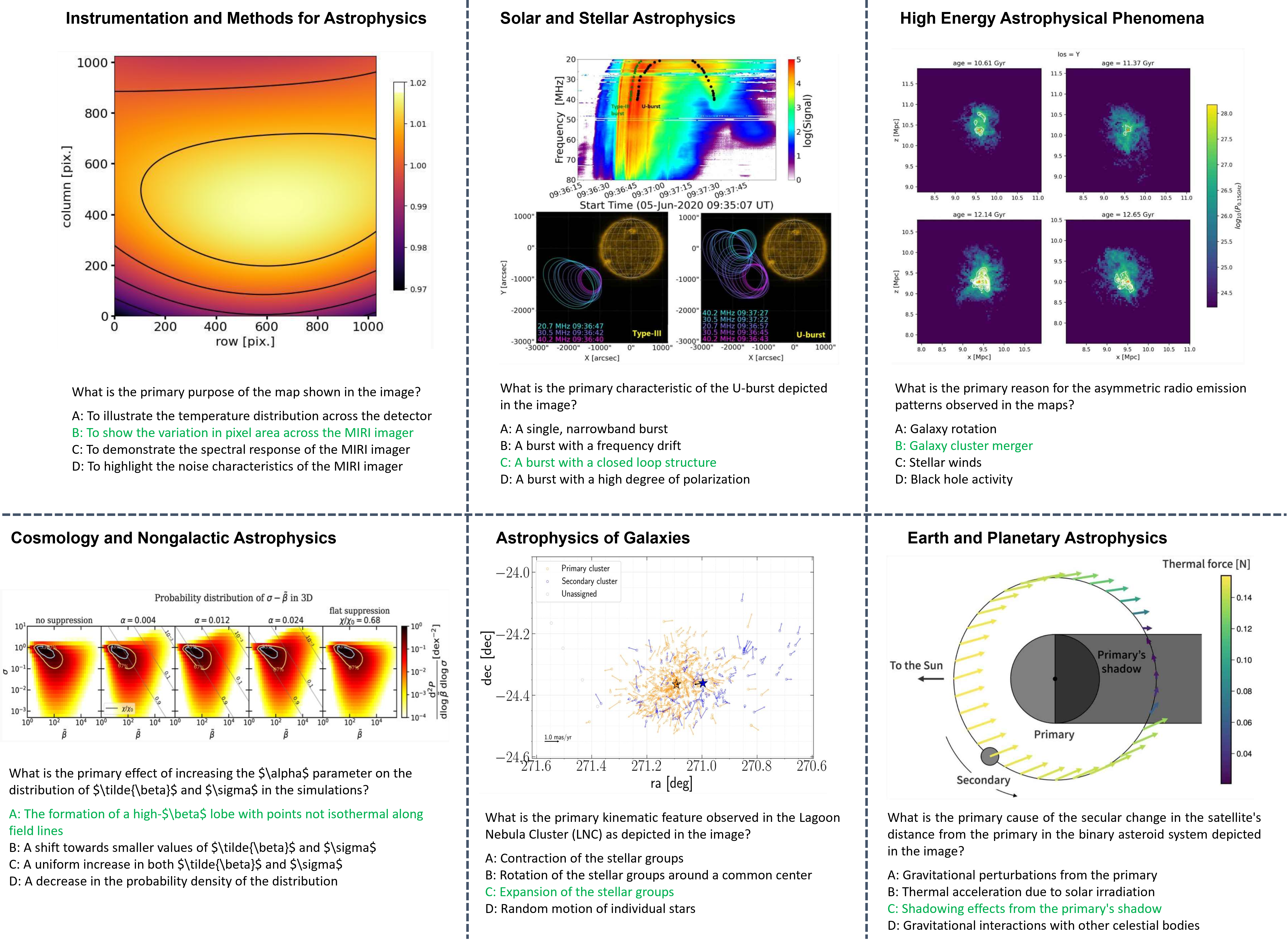}
    \caption{Examples of randomly selected questions in AstroMMBench.}
    \label{fig:samples}
\end{figure*}

Despite significant progress in MLLM evaluation across general and various specific domains, a dedicated multimodal evaluation benchmark specifically for astronomical images remains absent. Our work directly addresses this critical gap by introducing AstroMMBench.

\section{AstroMMBench} \label{sec:mmb}
\subsection{Overview Of AstroMMBench}
AstroMMBench is the first benchmark specifically designed to evaluate the performance of MLLMs in the domain of astronomical image interpretation. It comprises 621 multiple-choice questions, meticulously curated to cover six core subfields of astrophysics: Astrophysics of Galaxies (GA), Cosmology and Nongalactic Astrophysics (CO), Earth and Planetary Astrophysics (EP), High Energy Astrophysical Phenomena (HE), Instrumentation and Methods for Astrophysics (IM), and Solar and Stellar Astrophysics (SR). This structure ensures a broad yet deep assessment of MLLM performance across the discipline. 

As illustrated in Figure \ref{fig:categories-distribution}, the questions are well-distributed across these subfields, with counts ranging from 87 to 111 per category, ensuring representative topical coverage. Each question, paired with an astronomical image, requires a model to select the correct answer from four options. Figure \ref{fig:samples} showcases representative examples from AstroMMBench.

The construction of AstroMMBench involved a multi-stage process, detailed in the subsequent sections. This process began with the collection of image-text pairs from recent astrophysical literature (\S\ref{sec:data-collection}), followed by an automated pipeline for question generation (\S\ref{sec:questions-autogeneration}), and culminated in a rigorous, expert-led review phase to ensure the quality, relevance, and scientific accuracy of each question (\S\ref{sec:question-review}).

\subsection{Data Collection}\label{sec:data-collection}
Constructing a high-quality and domain-specific evaluation dataset that remains relevant in the rapidly evolving field of MLLMs presents challenges, particularly regarding data leakage. To address this, we need a data source that is both rich in domain-specific content and continuously updated. The arXiv repository perfectly fits this requirement, serving as a vast and dynamic archive of scientific preprints that reflect the very latest advancements across diverse subdisciplines of astrophysics. Its continuous nature allows for the potential generation of future benchmark versions utilizing data published after the training cutoff of new models, thereby mitigating the risk of data contamination.

For the initial construction of AstroMMBench, we focused exclusively on the "Astrophysics" (astro-ph) category on arXiv. We collected the TeX source files of 3,592 papers submitted between January 1, 2024, and July 31, 2024. From these collected papers, we extracted images along with their corresponding captions and contextual references found within the main body of the text, yielding an initial corpus of 19,299 image-text pairs.

This collection process, based on arXiv's constantly updating content, forms the foundation for a benchmark design that can be readily updated. While the specific timeframe of this initial dataset (collected in 2024) provides a snapshot of astrophysical research up to that point, the methodology allows for the creation of subsequent versions of AstroMMBench using newer arXiv data. This inherent flexibility is key to maintaining a high-quality benchmark that minimizes potential data leakage as MLLMs are continually trained on ever-larger and more recent datasets.

\subsection{Automatic Pipeline}\label{sec:automatic-pipline}

Manually creating high-quality exam questions, especially in specialized fields like astronomy, is not only time-consuming and laborious but may not be able to adapt to the ever-improving model development in a timely manner. With the rapid rise of MLLMs, it is possible to automatically generate high-quality questions from images with detailed text descriptions.

\begin{figure*}
    \centering
    \subfigure[Multiple Choice Questions Autogeneration]{
    \includegraphics[width=0.8\linewidth]{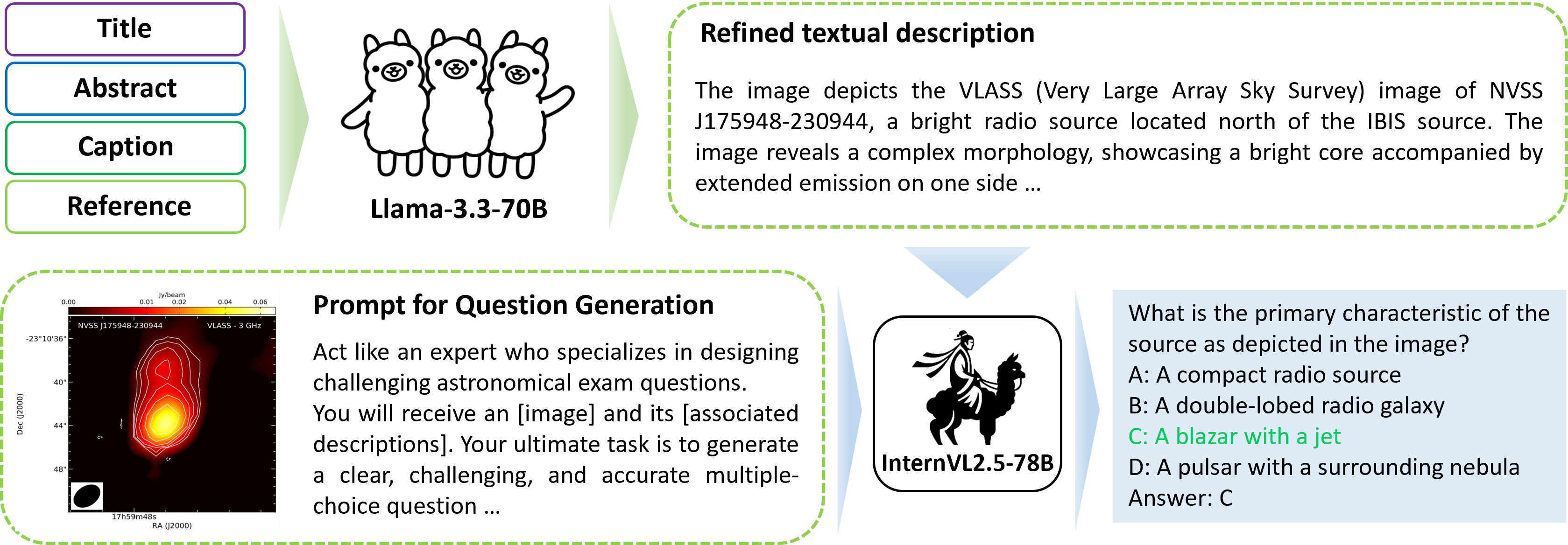}
    }
    \subfigure[Question/Answer Review]{\centering
    \includegraphics[width=0.8\linewidth]{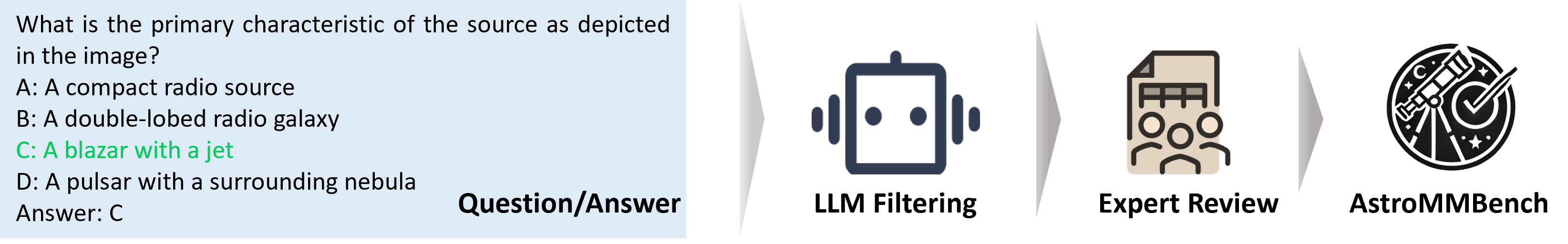}
    }
    \caption{Automated pipeline for multiple-choice question generation and review. The pipeline is divided into two stages. (a) The initial stage involves the autogeneration of multiple-choice questions. Llama-3.3-70B-Instruct refines textual descriptions associated with astronomical images, while InternVL2.5-78B generates corresponding questions. (b) The second stage is the review process, where the generated questions undergo filtering by large language models (LLMs) and expert evaluation to ensure the quality, correctness, and relevance of both the questions and answers before their inclusion in the final benchmark.}
    \label{fig:auto-pipeline}
\end{figure*}

To efficiently construct a large-scale benchmark, we developed an automated pipeline for question generation and curation. Specifically, we employed LLaMA3.3-70B-Instruct and InternVL2.5-78B \cite{chen2024internvl} for question generation. This automation significantly reduced the manual effort in building a large-scale benchmark. Figure \ref{fig:auto-pipeline} shows the entire automated process, which is divided into two main stages: stage one is used to generate multiple-choice questions, and stage two filters the generated questions through multi-step review to ensure question quality.

\subsubsection{Questions Autogeneration}\label{sec:questions-autogeneration}
The first stage of our pipeline focuses on automatically generating multiple-choice questions from the collected image-text pairs. Initially, we refine the textual data associated with each image to enhance its consistency and clarity. We observed that captions and contextual references extracted directly from research papers often contain:

\begin{itemize}
    \item \textbf{Information Redundancy}: Descriptions that include details irrelevant to the specific image or residual LaTeX formatting;
    \item \textbf{Style Inconsistency}: Variations in writing styles across different authors, which impact the standardization of the input text.
\end{itemize}
To address these challenges, we used the LLaMA3.3-70B-Instruct model to rewrite the textual data. This rewriting process was guided by a carefully designed prompt (see Appendix \ref{appendixA:prompt1}) aimed at ensuring the accuracy and completeness of the content while effectively reducing redundancy and unifying the expression style. To provide essential background context, we supplied the LLaMA model with the paper's title and abstract, in addition to the image captions and contextual references.

Following the text refinement, the polished textual descriptions, paired with their corresponding astronomical images, were input into the InternVL2.5-78B model to generate the multiple-choice questions. To ensure the generated questions were clear, challenging, and scientifically accurate, we utilized an implicit thought chain prompt (see Appendix \ref{appendixA:prompt2}). This prompt was designed to guide the model through a structured reasoning process, facilitating the generation of questions that effectively probe understanding of the image and its context, along with plausible answer options.
\subsubsection{Questions Review}\label{sec:question-review}

After generating preliminary questions, we introduced a multi-stage review process to ensure the quality and academic rigor of the generated questions. This is the core step of the second stage, which aims to ensure that the final retained questions are highly accurate and challenging through multi-model evaluation and expert review.

First, to ensure that the generated questions can effectively assess the respondents' ability to analyze astronomical images, we used five LLMs for an initial filtering step, including InternLM2.5-7B-Chat \cite{cai2024internlm2}, LLaMA-3.1-8B-Instruct, Yi-1.5-34B-Chat \cite{young2024yi}, Qwen2.5-32B \cite{qwen2.5}, and InternLM2-20B-Chat \cite{cai2024internlm2}. Each model answered each question five times, and a question was considered correctly answered by a model if at least three out of five responses were correct. To eliminate questions that could be answered primarily through linguistic reasoning without requiring visual understanding, we discarded questions that were correctly answered by at least two models. As a result of this filtering process, the initial pool of 19,299 generated questions was reduced to 9,677, retaining those more likely to require visual input for accurate interpretation.

\begin{figure}
    \centering
    \includegraphics[width=0.8\linewidth]{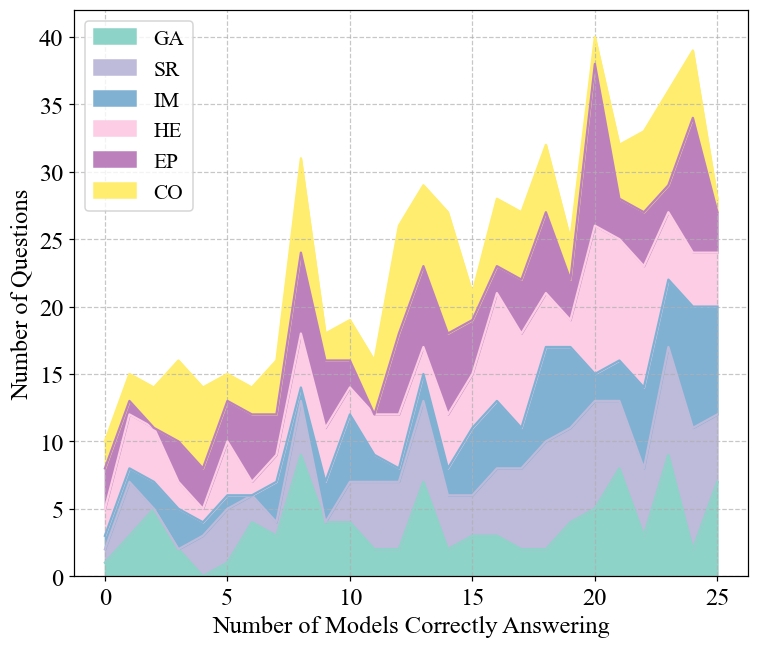}
    \caption{Distribution of question difficulty in AstroMMBench, based on the number of evaluated models that correctly answered each question. The x-axis indicates the "Number of Models Correctly Answering" a question (0-25), and the y-axis shows the count of questions at each correctness level, broken down by subfield.}
    \label{fig:difficulty}
\end{figure}

To ensure the accuracy, relevance, and rigor of the dataset, a panel of 15 astronomy experts—each holding at least a master's degree in astronomy or a related field—conducted a thorough review of 1,800 randomly selected questions from the remaining 9,677 questions. Each question was independently evaluated by an expert within the corresponding subfield. Based on criteria including image-question alignment, contextual completeness, answer accuracy and uniqueness, and the necessity of domain-specific knowledge, a total of 621 high-quality questions were retained. Furthermore, to mitigate potential biases in model responses and ensure fair evaluation, the correct answer options for these 621 questions were reassigned to be uniformly distributed across options A, B, C, and D.

\subsection{Difficulty Distribution}\label{sec:difficulty-distribution}

To further characterize AstroMMBench as an evaluation benchmark, we analyzed the distribution of question difficulty. The difficulty of each question in AstroMMBench, as revealed by the evaluation results presented in Section \ref{sec:experiments}, is characterized by the number of the 25 evaluated models that were able to correctly answer it. Figure \ref{fig:difficulty} illustrates this difficulty distribution based on the performance of the evaluated models. The x-axis represents the number of models that correctly answered a given question, ranging from 0 for the most challenging questions (answered correctly by no models) to 25 for the easiest (answered correctly by all models). The y-axis shows the number of questions at each level of correctness, with stacked areas representing the contribution of each astrophysical subfield.

The overall trend in the figure indicates that the number of questions gradually decreases as their difficulty increases. This suggests that the benchmark deliberately avoids an overrepresentation of extremely difficult questions that current models struggle to solve. Most questions fall within the medium difficulty range, which is effective for differentiating model capabilities. Additionally, the stacked area chart shows that questions from all subfields contribute to the overall difficulty spectrum, although their proportions vary across difficulty levels. This distribution demonstrates that AstroMMBench offers a challenging yet balanced evaluation framework.

\section{Experiments} \label{sec:experiments}

\begin{table*}
\small
\centering
\setlength{\tabcolsep}{4pt}
\caption{Performance of 3 closed-source and 22 open-source models on the AstroMMBench dataset across sub-domains of astrophysics. The best-performing model in each category is in bold.}
\begin{tabular}{lcccccccc}
\hline
\textbf{Model}                             & \textbf{\begin{tabular}[c]{@{}c@{}}Overall\\ (621)\end{tabular}} & \textbf{\begin{tabular}[c]{@{}c@{}}GA\\ (97)\end{tabular}} & \textbf{\begin{tabular}[c]{@{}c@{}}CO\\ (111)\end{tabular}} & \textbf{\begin{tabular}[c]{@{}c@{}}EP\\ (105)\end{tabular}} & \textbf{\begin{tabular}[c]{@{}c@{}}HE\\ (110)\end{tabular}} & \textbf{\begin{tabular}[c]{@{}c@{}}IM\\ (87)\end{tabular}} & \textbf{\begin{tabular}[c]{@{}c@{}}SR\\ (111)\end{tabular}} & \textbf{OpenCompass} \\ \hline
\multicolumn{9}{c}{\textit{\textbf{Closed-source Models}}}                                                                                                       \\
\hline
\multicolumn{1}{l|}{ChatGPT-4o \cite{hurst2024gpt}}                 & 69.07                                                           & 69.07                                                     & {\textbf{67.57}}                                       & 67.62                                                      & 70.91                                                      & 68.97                                                     & \multicolumn{1}{c|}{71.17}                                 & {\textbf{47.49}} \\
\multicolumn{1}{l|}{Doubao-1.5-vision-pro} & 68.12                                                           & {\textbf{70.10}}                                      & {\textbf{67.57}}                                       & 64.76                                                      & {\textbf{72.73}}                                       & 67.82                                                     & \multicolumn{1}{c|}{65.77}                                 & --                   \\
\multicolumn{1}{l|}{QwenVLMax \cite{Qwen-VL}}             & 66.83                                                           & 58.76                                                     & 58.56                                                      & 69.52                                                      & 65.45                                                      & 78.16                                                     & \multicolumn{1}{c|}{72.07}                                 & --                   \\ \hline
\multicolumn{9}{c}{\textit{\textbf{Open-source Models}}}                                                                                  \\ \hline
\multicolumn{1}{l|}{Ovis2-34B \cite{lu2024ovis}}             & {\textbf{70.53}}                                            & 68.04                                                     & {\textbf{67.57}}                                       & 68.57                                                      & {\textbf{72.73}}                                       & 78.16                                                     & \multicolumn{1}{c|}{69.37}                                 & 42.82                \\
\multicolumn{1}{l|}{InternVL3-38B \cite{zhu2025internvl3}}         & 67.63                                                           & 68.04                                                     & 53.15                                                      & 61.90                                                      & 70.91                                                      & {\textbf{80.46}}                                      & \multicolumn{1}{c|}{{\textbf{73.87}}}                  & 45.55                \\
\multicolumn{1}{l|}{Qwen2.5-VL-72B \cite{bai2025qwen2}}        & 67.47                                                           & 59.79                                                     & 57.66                                                      & {\textbf{72.38}}                                       & 69.09                                                      & 74.71                                                     & \multicolumn{1}{c|}{72.07}                                 & 48.25                \\
\multicolumn{1}{l|}{Ovis2-16B \cite{lu2024ovis}}             & 67.31                                                           & 63.92                                                     & 63.06                                                      & 61.90                                                      & 70.91                                                      & 75.86                                                     & \multicolumn{1}{c|}{69.37}                                 & 39.60                \\
\multicolumn{1}{l|}{Qwen2.5-VL-32B \cite{bai2025qwen2}}        & 64.25                                                           & 57.73                                                     & 60.36                                                      & 60.00                                                      & 68.18                                                      & 68.97                                                     & \multicolumn{1}{c|}{70.27}                                 & --                   \\
\multicolumn{1}{l|}{InternVL3-78B \cite{zhu2025internvl3}}         & 64.25                                                           & 63.92                                                     & 51.35                                                      & 60.95                                                      & 65.45                                                      & 73.56                                                     & \multicolumn{1}{c|}{72.07}                                 & 45.96                \\
\multicolumn{1}{l|}{InternVL3-14B \cite{zhu2025internvl3}}         & 63.77                                                           & 61.86                                                     & 53.15                                                      & 62.86                                                      & 61.82                                                      & 73.56                                                     & \multicolumn{1}{c|}{71.17}                                 & 40.72                \\
\multicolumn{1}{l|}{SAIL-VL-1.6-8B \cite{dong2025scalable}}        & 62.32                                                           & 56.70                                                     & 58.56                                                      & 63.81                                                      & 62.73                                                      & 68.97                                                     & \multicolumn{1}{c|}{63.96}                                 & 37.92                \\
\multicolumn{1}{l|}{InternVL3-8B \cite{zhu2025internvl3}}          & 61.03                                                           & 59.79                                                     & 52.25                                                      & 61.90                                                      & 60.91                                                      & 66.67                                                     & \multicolumn{1}{c|}{65.77}                                 & 37.40                \\
\multicolumn{1}{l|}{InternVL3-9B \cite{zhu2025internvl3}}          & 60.55                                                           & 60.82                                                     & 49.55                                                      & 55.24                                                      & 62.73                                                      & 65.52                                                     & \multicolumn{1}{c|}{70.27}                                 & --                   \\
\multicolumn{1}{l|}{DeepSeek\_VL2 \cite{wu2024deepseekvl2}}         & 59.90                                                           & 57.73                                                     & 53.15                                                      & 61.90                                                      & 61.82                                                      & 68.97                                                     & \multicolumn{1}{c|}{57.66}                                 & 38.70                \\
\multicolumn{1}{l|}{Qwen2.5-VL-3B \cite{bai2025qwen2}}         & 58.94                                                           & 52.58                                                     & 59.46                                                      & 56.19                                                      & 58.18                                                      & 65.52                                                     & \multicolumn{1}{c|}{62.16}                                 & 38.33                \\
\multicolumn{1}{l|}{MiniCPM-o-2.6 \cite{yao2024minicpm}}         & 57.97                                                           & 54.64                                                     & 51.35                                                      & 50.48                                                      & 62.73                                                      & 67.82                                                     & \multicolumn{1}{c|}{62.16}                                 & 34.67                \\
\multicolumn{1}{l|}{Qwen2.5-VL-7B \cite{bai2025qwen2}}         & 57.33                                                           & 52.58                                                     & 56.76                                                      & 53.33                                                      & 54.55                                                      & 67.82                                                     & \multicolumn{1}{c|}{60.36}                                 & 43.21                \\
\multicolumn{1}{l|}{Kimi-VL-A3B-Instruct \cite{kimiteam2025kimivl}}  & 56.68                                                           & 50.52                                                     & 54.95                                                      & 54.29                                                      & 55.45                                                      & 68.97                                                     & \multicolumn{1}{c|}{57.66}                                 & 37.00                \\
\multicolumn{1}{l|}{LLaVA\_Onevision\_72B \cite{li2024llava}} & 55.39                                                           & 52.58                                                     & 54.95                                                      & 53.33                                                      & 50.91                                                      & 63.22                                                     & \multicolumn{1}{c|}{58.56}                                 & 39.05                \\
\multicolumn{1}{l|}{Gemma3-12B \cite{gemma_2025}}            & 52.82                                                           & 49.48                                                     & 51.35                                                      & 60.95                                                      & 45.45                                                      & 50.57                                                     & \multicolumn{1}{c|}{58.56}                                 & 34.15                \\
\multicolumn{1}{l|}{InternVL3-2B \cite{zhu2025internvl3}}          & 51.69                                                           & 53.61                                                     & 47.75                                                      & 46.67                                                      & 50.00                                                      & 55.17                                                     & \multicolumn{1}{c|}{57.66}                                 & 30.96                \\
\multicolumn{1}{l|}{Kimi-VL-A3B-Thinking \cite{kimiteam2025kimivl}}  & 50.08                                                           & 50.52                                                     & 45.05                                                      & 43.81                                                      & 47.27                                                      & 57.47                                                     & \multicolumn{1}{c|}{57.66}                                 & --                   \\
\multicolumn{1}{l|}{GLM-4v-9B \cite{glm2024chatglm}}             & 49.76                                                           & 46.39                                                     & 39.64                                                      & 48.57                                                      & 52.73                                                      & 59.77                                                     & \multicolumn{1}{c|}{53.15}                                 & 37.85                \\
\multicolumn{1}{l|}{InternVL3-1B \cite{zhu2025internvl3}}          & 45.73                                                           & 47.42                                                     & 38.74                                                      & 39.05                                                      & 42.73                                                      & 57.47                                                     & \multicolumn{1}{c|}{51.35}                                 & 24.39                \\
\multicolumn{1}{l|}{Gemma3-4B \cite{gemma_2025}}             & 42.51                                                           & 42.27                                                     & 34.23                                                      & 39.05                                                      & 41.82                                                      & 48.28                                                     & \multicolumn{1}{c|}{50.45}                                 & 32.21                \\ \hline
\end{tabular}
\label{table:results}
\end{table*}

\subsection{Baselines}\label{sec:baselines}
\paragraph{MLLMs} To evaluate the performance of current MLLMs in the domain of astronomy, we selected a diverse set of 25 models, comprising 22 publicly available open-source models and 3 carefully selected powerful closed-source models. The complete list of evaluated models, along with their overall and subfield-specific performance on AstroMMBench, is presented in Table \ref{table:results}.

\paragraph{Evaluation} For the evaluation process, we utilized VLMEvalKit \cite{duan2024vlmevalkit}, a widely used open-source evaluation framework specifically designed for MLLMs, which provides standardized protocols and metrics. As AstroMMBench is composed exclusively of multiple-choice questions with a single correct answer, the primary evaluation metric used is accuracy (proportion of correctly answered questions). A model's response is considered correct only if the extracted answer option precisely matches the predefined correct answer. To accurately extract the chosen answer option from the potentially verbose text outputs of the evaluated MLLMs, we employed DeepSeek-V3 \cite{deepseekai2024deepseekv3} to parse model responses and identify the intended answer option (A, B, C, or D), thereby mitigating issues arising from simple pattern matching in free-form generation. 
All experiments were conducted on hardware equipped with eight NVIDIA A100 GPUs.

{\subsection{Main Results on AstroMMBench}}
\subsubsection{Overall Performance}
Table \ref{table:results} summarizes the performance of 25 MLLMs evaluated on AstroMMBench, sorted by overall accuracy. The OpenCompass scores are drawn from the OpenCompass multimodal model leaderboard\footnote{\url{https://rank.opencompass.org.cn/leaderboard-multimodal/?m=REALTIME}}, which reflects model capabilities across general-purpose tasks. The results demonstrate substantial variation in performance across models. Among them, the open-source Ovis2-34B model achieved the highest overall accuracy (70.53\%), outperforming all other models on this benchmark. It is followed by ChatGPT-4o (69.07\%) and Doubao-1.5-Vision-Pro (68.12\%), highlighting the competitiveness of state-of-the-art commercial MLLMs. Remarkably, Ovis2-34B’s leading performance over these proprietary models underscores the rapid advancement and potential of open-source MLLMs for domain-specific tasks.

The scores for other models span a wide range, with Gemma3-4B and InternVL3-1B achieving the lowest overall accuracies of 42.51\% and 45.73\%, respectively. Although all models outperform the 25\% accuracy expected from random guessing on a four-choice multiple-choice task, their overall performance remains limited. This highlights the difficulty of the AstroMMBench benchmark and reveals significant room for improvement in current MLLMs’ ability to process astronomical images.

\subsubsection{Relationship with General Capabilities}\label{sec:general-correlation}
\begin{figure*}
    \centering
    \includegraphics[width=0.77\linewidth]{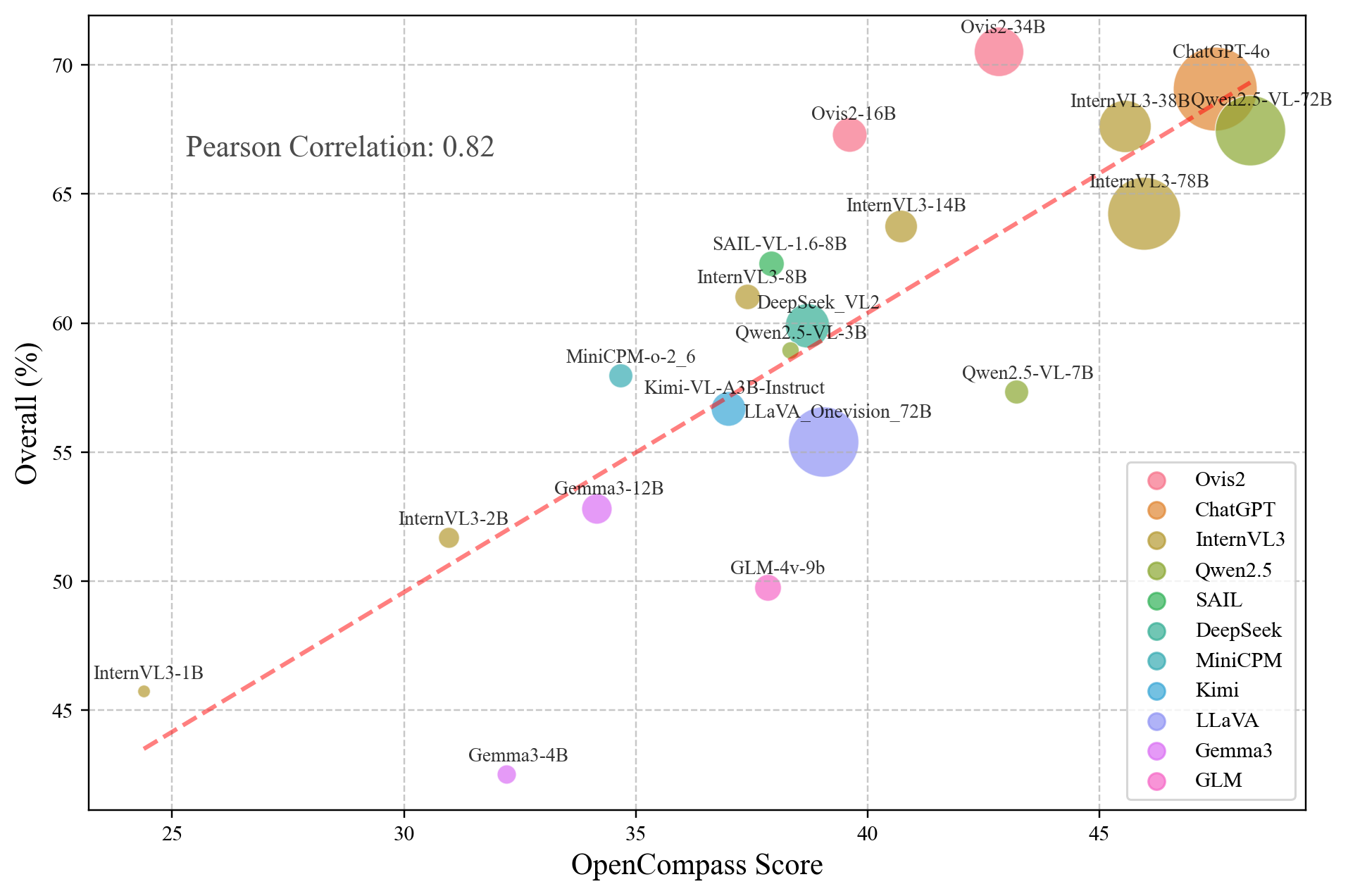}
    \caption{Relationship between general multimodal performance (OpenCompass score) and specialized astronomical image interpretation performance (AstroMMBench overall accuracy) for 22 MLLMs. Point size represents model scale (parameter count)}
    \label{fig:general-overall}
\end{figure*}

As illustrated in Figure \ref{fig:general-overall}, there is a clear positive correlation between the models' general performance (OpenCompass score) and their astronomical domain performance (AstroMMBench overall score). The red dashed line in the figure represents the linear regression fit between the two, revealing a linear correlation. To quantify the strength of this linear relationship, we calculated the Pearson correlation coefficient, obtaining $(r = 0.82 )$, which demonstrates a significant positive correlation. The calculation method for the Pearson correlation coefficient $(r)$ is provided in equation \ref{eq:pearson}. This suggests that models performing well on general tasks also tend to excel in astrophysical tasks, validating the robustness and scientific soundness of AstroMMBench.

\begin{equation}
r = \frac{\sum (X - \bar{X})(Y - \bar{Y})}{\sqrt{\sum (X - \bar{X})^2 \sum (Y - \bar{Y})^2}}
\label{eq:pearson}
\end{equation}


However, this correlation is not without exceptions. For example, Ovis2-34B outperforms models with higher general scores like ChatGPT-4o, Qwen2.5-VL-72B, and InternVL3-38B on AstroMMBench. This anomaly suggests that while general multimodal capabilities can predict success in specialized fields, some models may face difficulties when confronted with domain-specific challenges in astrophysics. It also underscores the unique challenges posed by AstroMMBench, where models must handle domain-specific questions that may not be adequately captured by general-purpose multimodal benchmarks.

\subsubsection{Analysis by Subfield}\label{sec:subfield-analysis}
\begin{figure}
    \centering
    \includegraphics[width=0.8\linewidth]{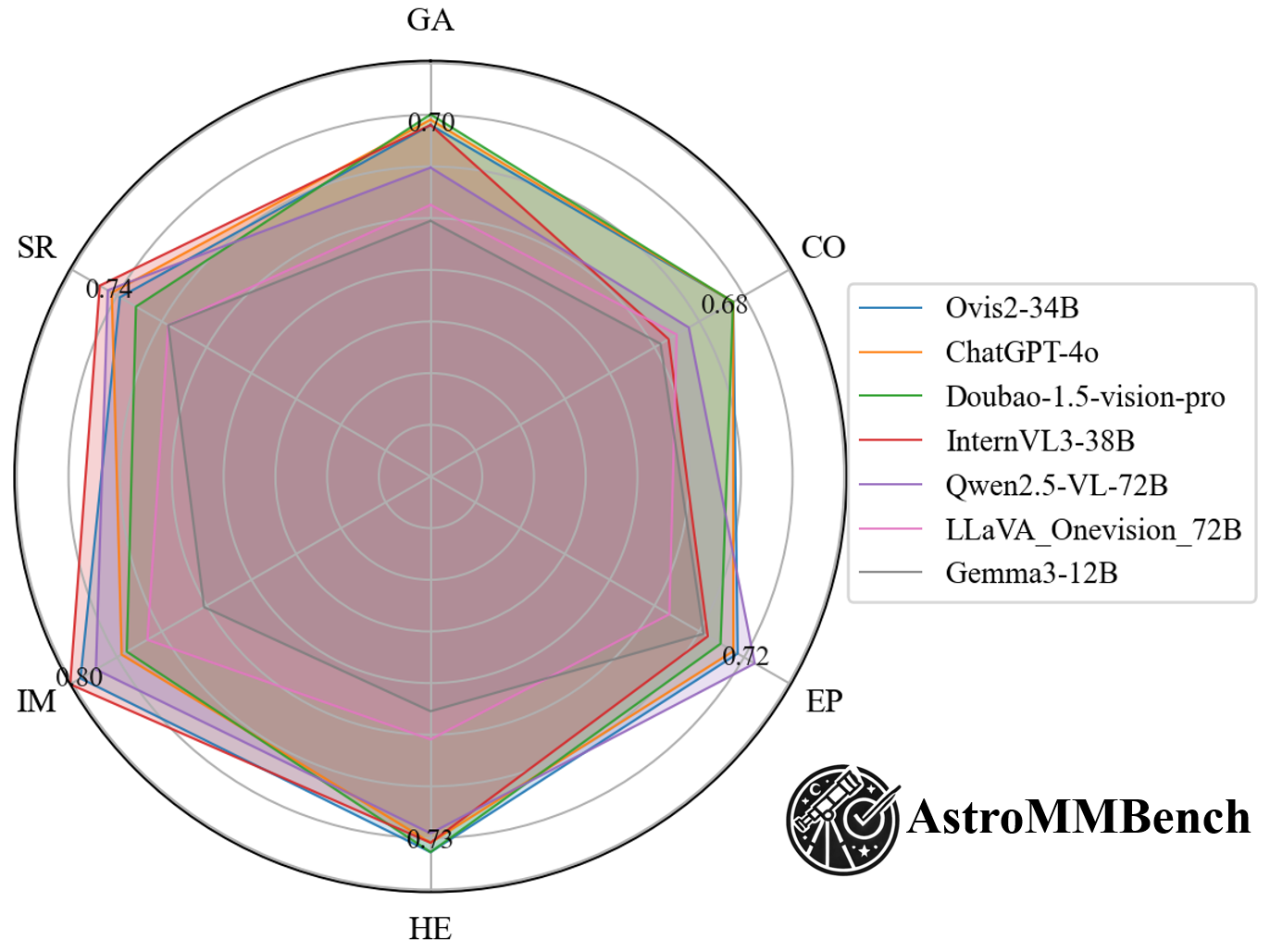}
    \caption{Comparison of model performance across six astrophysical subfields in AstroMMBench.}
    \label{fig:subfields}
\end{figure}

AstroMMBench encompasses six major subfields of astrophysics. A detailed analysis of model performance within these distinct domains allows for a deeper understanding of their strengths and potential limitations when tackling different types of astronomical tasks. Figure \ref{fig:subfields} presents a radar chart offering a visual overview of the performance profiles for selected representative models across the six subfields. Examples of these models' responses within each subfield and varying difficulty questions are provided in Appendix \ref{appendixB}.

Our analysis indicates that performance disparities across different subfields reflect the varying capabilities required by the questions in each category. Specifically, questions in the IM and SR subfields primarily demand skills related to interpreting standard astronomical plots (e.g., time series, relationships between physical quantities) and recognizing common astronomical objects or instrument components. These tasks may align well with the graph understanding and object recognition capabilities models acquire during general domain training, thus resulting in generally higher scores in these subfields. Conversely, questions in the CO and HE subfields typically require a deeper understanding of abstract theoretical concepts, interpretation of highly specialized or unconventional visualizations (e.g., statistical maps of cosmic structures, signatures of particle interactions), and complex multi-step reasoning based on fragmented information. These capabilities may be less developed or consistently present in current general-purpose MLLMs. The GA and EP subfields, covering a wide range of question types from galaxy morphological classification to interpreting planetary atmospheric data or orbital dynamics plots, require a mix of these abilities and exhibit intermediate difficulty.

The radar chart in Figure \ref{fig:subfields} visualizes the performance profiles of various models across astrophysical subfields. Top performers like Ovis2-34B and ChatGPT-4o display balanced, consistent polygons, underscoring their robustness and versatility in astronomy. In contrast, InternVL3-38B, while achieving leading scores in the IM and SR fields, shows a notable decrease in performance in the CO field. This suggests that its ability to interpret standard astronomical plots might be stronger than its capacity to handle the more abstract concepts and specialized imagery common in cosmology. Other models also showcase their specific characteristics, such as Doubao-1.5-vision-pro's prominent strength in the HE field and Qwen2.5-VL-72B's leading performance in the EP field. These variations highlight that different models may possess specific proficiencies aligned with particular astrophysical domains, likely stemming from differences in their training or architecture.

\section{Conclusion} \label{sec:conclusion}
In this paper, we introduce AstroMMBench, the first benchmark tailored to assess MLLMs in astronomy. It features 621 multiple-choice questions spanning six key astrophysics subfields, automatically generated and expert-reviewed for accuracy and relevance.

Using the VLMEvalKit framework, we evaluated 25 MLLMs and observed significant performance differences. The open-source Ovis2-34B outperformed top closed-source models like ChatGPT-4o and Doubao-1.5-vision-pro with a 70.53\% score, emphasizing the promise of open models in scientific domains. We found a strong positive correlation between general MLLM performance and AstroMMBench scores, yet exceptions demonstrate the critical need for domain-specific evaluation to truly assess specialized proficiency. Furthermore, performance varied across astrophysical subfields, with domains like Cosmology and Nongalactic Astrophysics and High Energy Astrophysics proving generally more challenging than Instrumentation and Methods for Astrophysics and Solar and Stellar Astrophysics, reflecting the diverse demands on MLLM capabilities. We hope that AstroMMBench can become a continuously evolving platform to support the evaluation and promotion of the next generation of MLLMs in astronomy.

\section*{Limitations}
Our study provides a benchmark and framework for the performance of multimodal large language models in the astronomy domain, but we acknowledge that there are several limitations to our study that may require further exploration:

\paragraph{Limited benchmark size and task diversity} The current benchmark size of 621 questions, while substantial for a first benchmark of this nature, is relatively limited compared to the vastness and complexity of astronomical phenomena and tasks. Furthermore, the task format is currently restricted to multiple-choice Visual Question Answering. Incorporating more diverse question types, such as open-ended questions, multi-step reasoning tasks, or predictive analysis challenges, would provide a more comprehensive assessment of MLLMs' advanced capabilities needed for complex scientific analysis beyond direct VQA.

\paragraph{Challenges in automated question generation and curation} Although an automated pipeline is employed for initial question generation from scientific literature, the quality of the generated questions can be inconsistent. This often results in a proportion of low-quality or irrelevant questions that do not adequately test specialized astronomical knowledge. Consequently, ensuring the scientific rigor and quality of the final benchmark set heavily relies on a costly and time-consuming manual expert review process. This reliance on manual curation limits the scalability and efficiency of expanding the benchmark size and providing frequent updates with new data, presenting a key challenge for maintaining a dynamic benchmark.


In our future work, we will focus on addressing these shortcomings to overall improve the quality and scalability of AstroMMBench while ensuring that it becomes a comprehensive and evolving benchmark for evaluating MLLMs in the specialized field of astronomy.

\section*{Ethics Statement}
\paragraph{Copyright and License} Regarding the data used in AstroMMBench, all images and associated textual content are sourced from publicly available preprints on arXiv. We ensure compliance with copyright regulations by strictly adhering to the terms of use for arXiv data, which permits the re-use and distribution of content under specific licenses (typically Creative Commons licenses as specified by the authors). We maintain adherence to the established legal and ethical standards for using publicly available scientific literature.

We are committed to making AstroMMBench openly accessible to the research community to facilitate further research and evaluation of MLLMs in astronomy. AstroMMBench will be released under the Creative Commons Attribution 4.0 International License (CC BY 4.0). 

\bibliography{custom}

\begin{thebibliography}{55}
\expandafter\ifx\csname natexlab\endcsname\relax\def\natexlab#1{#1}\fi

\bibitem[{Abdin et~al.(2024)Abdin, Aneja, Awadalla, Awadallah, Awan, Bach, Bahree, Bakhtiari, Bao, Behl et~al.}]{Abdin2024Phi3TR}
Marah Abdin, Jyoti Aneja, Hany Awadalla, Ahmed Awadallah, Ammar~Ahmad Awan, Nguyen Bach, Amit Bahree, Arash Bakhtiari, Jianmin Bao, Harkirat Behl, et~al. 2024.
\newblock Phi-3 technical report: A highly capable language model locally on your phone.
\newblock \emph{arXiv preprint arXiv:2404.14219}.

\bibitem[{Achiam et~al.(2023)Achiam, Adler, Agarwal, Ahmad, Akkaya, Aleman, Almeida, Altenschmidt, Altman, Anadkat et~al.}]{Achiam2023GPT4TR}
Josh Achiam, Steven Adler, Sandhini Agarwal, Lama Ahmad, Ilge Akkaya, Florencia~Leoni Aleman, Diogo Almeida, Janko Altenschmidt, Sam Altman, Shyamal Anadkat, et~al. 2023.
\newblock Gpt-4 technical report.
\newblock \emph{arXiv preprint arXiv:2303.08774}.

\bibitem[{Bai et~al.(2023{\natexlab{a}})Bai, Bai, Chu, Cui, Dang, Deng, Fan, Ge, Han, Huang, Hui, Ji, Li, Lin, Lin, Liu, Liu, Lu, Lu, Ma, Men, Ren, Ren, Tan, Tan, Tu, Wang, Wang, Wang, Wu, Xu, Xu, Yang, Yang, Yang, Yang, Yang, Yao, Yu, Bowen, Yuan, Yuan, Zhang, Zhang, Zhang, Zhang, Zhou, Zhou, Zhou, and Zhu}]{Bai2023QwenTR}
Jinze Bai, Shuai Bai, Yunfei Chu, Zeyu Cui, Kai Dang, Xiaodong Deng, Yang Fan, Wenhang Ge, Yu~Han, Fei Huang, Binyuan Hui, Luo Ji, Mei Li, Junyang Lin, Runji Lin, Dayiheng Liu, Gao Liu, Chengqiang Lu, K.~Lu, Jianxin Ma, Rui Men, Xingzhang Ren, Xuancheng Ren, Chuanqi Tan, Sinan Tan, Jianhong Tu, Peng Wang, Shijie Wang, Wei Wang, Shengguang Wu, Benfeng Xu, Jin Xu, An~Yang, Hao Yang, Jian Yang, Jian Yang, Shusheng Yang, Yang Yao, Bowen Yu, Yu~Bowen, Hongyi Yuan, Zheng Yuan, Jianwei Zhang, Xing Zhang, Yichang Zhang, Zhenru Zhang, Chang Zhou, Jingren Zhou, Xiaohuan Zhou, and Tianhang Zhu. 2023{\natexlab{a}}.
\newblock \href {https://api.semanticscholar.org/CorpusID:263134555} {Qwen technical report}.
\newblock \emph{ArXiv}, abs/2309.16609.

\bibitem[{Bai et~al.(2023{\natexlab{b}})Bai, Bai, Yang, Wang, Tan, Wang, Lin, Zhou, and Zhou}]{Qwen-VL}
Jinze Bai, Shuai Bai, Shusheng Yang, Shijie Wang, Sinan Tan, Peng Wang, Junyang Lin, Chang Zhou, and Jingren Zhou. 2023{\natexlab{b}}.
\newblock Qwen-vl: A versatile vision-language model for understanding, localization, text reading, and beyond.
\newblock \emph{arXiv preprint arXiv:2308.12966}.

\bibitem[{Bai et~al.(2025)Bai, Chen, Liu, Wang, Ge, Song, Dang, Wang, Wang, Tang et~al.}]{bai2025qwen2}
Shuai Bai, Keqin Chen, Xuejing Liu, Jialin Wang, Wenbin Ge, Sibo Song, Kai Dang, Peng Wang, Shijie Wang, Jun Tang, et~al. 2025.
\newblock Qwen2. 5-vl technical report.
\newblock \emph{arXiv preprint arXiv:2502.13923}.

\bibitem[{Brown et~al.(2020)Brown, Mann, Ryder, Subbiah, Kaplan, Dhariwal, Neelakantan, Shyam, Sastry, Askell et~al.}]{Brown2020LanguageMA}
Tom Brown, Benjamin Mann, Nick Ryder, Melanie Subbiah, Jared~D Kaplan, Prafulla Dhariwal, Arvind Neelakantan, Pranav Shyam, Girish Sastry, Amanda Askell, et~al. 2020.
\newblock Language models are few-shot learners.
\newblock \emph{Advances in neural information processing systems}, 33:1877--1901.

\bibitem[{Cai et~al.(2024)Cai, Cao, Chen, Chen, Chen, Chen, Chen, Chen, Chen, Chu et~al.}]{cai2024internlm2}
Zheng Cai, Maosong Cao, Haojiong Chen, Kai Chen, Keyu Chen, Xin Chen, Xun Chen, Zehui Chen, Zhi Chen, Pei Chu, et~al. 2024.
\newblock Internlm2 technical report.
\newblock \emph{arXiv preprint arXiv:2403.17297}.

\bibitem[{Chen et~al.(2024{\natexlab{a}})Chen, Li, Dong, Zhang, Zang, Chen, Duan, Wang, Qiao, Lin et~al.}]{chen2024mmstar}
Lin Chen, Jinsong Li, Xiaoyi Dong, Pan Zhang, Yuhang Zang, Zehui Chen, Haodong Duan, Jiaqi Wang, Yu~Qiao, Dahua Lin, et~al. 2024{\natexlab{a}}.
\newblock Are we on the right way for evaluating large vision-language models?
\newblock \emph{arXiv preprint arXiv:2403.20330}.

\bibitem[{Chen et~al.(2024{\natexlab{b}})Chen, Wang, Cao, Liu, Gao, Cui, Zhu, Ye, Tian, Liu et~al.}]{chen2024expanding}
Zhe Chen, Weiyun Wang, Yue Cao, Yangzhou Liu, Zhangwei Gao, Erfei Cui, Jinguo Zhu, Shenglong Ye, Hao Tian, Zhaoyang Liu, et~al. 2024{\natexlab{b}}.
\newblock Expanding performance boundaries of open-source multimodal models with model, data, and test-time scaling.
\newblock \emph{arXiv preprint arXiv:2412.05271}.

\bibitem[{Chen et~al.(2024{\natexlab{c}})Chen, Wang, Tian, Ye, Gao, Cui, Tong, Hu, Luo, Ma et~al.}]{chen2024internvl}
Zhe Chen, Weiyun Wang, Hao Tian, Shenglong Ye, Zhangwei Gao, Erfei Cui, Wenwen Tong, Kongzhi Hu, Jiapeng Luo, Zheng Ma, et~al. 2024{\natexlab{c}}.
\newblock How far are we to gpt-4v? closing the gap to commercial multimodal models with open-source suites.
\newblock \emph{arXiv preprint arXiv:2404.16821}.

\bibitem[{Cui et~al.(2023)Cui, Ma, Cao, Ye, Zhou, Liang, Chen, Lu, Yang, Liao, Gao, Li, Tang, Cao, Zhou, Liu, Yan, Mei, Cao, Wang, and Zheng}]{cui2023autodrive}
Can Cui, Yunsheng Ma, Xu~Cao, Wenqian Ye, Yang Zhou, Kaizhao Liang, Jintai Chen, Juanwu Lu, Zichong Yang, Kuei-Da Liao, Tianren Gao, Erlong Li, Kun Tang, Zhipeng Cao, Tongxi Zhou, Ao~Liu, Xinrui Yan, Shuqi Mei, Jianguo Cao, Ziran Wang, and Chao Zheng. 2023.
\newblock \href {https://api.semanticscholar.org/CorpusID:265308931} {A survey on multimodal large language models for autonomous driving}.
\newblock \emph{2024 IEEE/CVF Winter Conference on Applications of Computer Vision Workshops (WACVW)}, pages 958--979.

\bibitem[{DeepSeek-AI(2024)}]{deepseekai2024deepseekv3}
DeepSeek-AI. 2024.
\newblock \href {http://arxiv.org/abs/2412.19437} {Deepseek-v3 technical report}.

\bibitem[{Devlin(2018)}]{devlin2018bert}
Jacob Devlin. 2018.
\newblock Bert: Pre-training of deep bidirectional transformers for language understanding.
\newblock \emph{arXiv preprint arXiv:1810.04805}.

\bibitem[{Dong et~al.(2025)Dong, Kang, Yin, Liang, Feng, and Ran}]{dong2025scalable}
Hongyuan Dong, Zijian Kang, Weijie Yin, Xiao Liang, Chao Feng, and Jiao Ran. 2025.
\newblock Scalable vision language model training via high quality data curation.
\newblock \emph{arXiv preprint arXiv:2501.05952}.

\bibitem[{Duan et~al.(2024)Duan, Yang, Qiao, Fang, Chen, Liu, Dong, Zang, Zhang, Wang et~al.}]{duan2024vlmevalkit}
Haodong Duan, Junming Yang, Yuxuan Qiao, Xinyu Fang, Lin Chen, Yuan Liu, Xiaoyi Dong, Yuhang Zang, Pan Zhang, Jiaqi Wang, et~al. 2024.
\newblock Vlmevalkit: An open-source toolkit for evaluating large multi-modality models.
\newblock In \emph{Proceedings of the 32nd ACM International Conference on Multimedia}, pages 11198--11201.

\bibitem[{Fini et~al.(2024)Fini, Shukor, Li, Dufter, Klein, Haldimann, Aitharaju, da~Costa, Béthune, Gan, Toshev, Eichner, Nabi, Yang, Susskind, and El-Nouby}]{fini2024AIMv2e}
Enrico Fini, Mustafa Shukor, Xiujun Li, Philipp Dufter, Michal Klein, David Haldimann, Sai Aitharaju, Victor Guilherme~Turrisi da~Costa, Louis Béthune, Zhe Gan, Alexander~T Toshev, Marcin Eichner, Moin Nabi, Yinfei Yang, Joshua~M. Susskind, and Alaaeldin El-Nouby. 2024.
\newblock \href {http://arxiv.org/abs/2411.14402} {Multimodal autoregressive pre-training of large vision encoders}.

\bibitem[{Fu et~al.(2024)Fu, Chen, Shen, Qin, Zhang, Lin, Yang, Zheng, Li, Sun, Wu, and Ji}]{fu2024mme}
Chaoyou Fu, Peixian Chen, Yunhang Shen, Yulei Qin, Mengdan Zhang, Xu~Lin, Jinrui Yang, Xiawu Zheng, Ke~Li, Xing Sun, Yunsheng Wu, and Rongrong Ji. 2024.
\newblock \href {http://arxiv.org/abs/2306.13394} {Mme: A comprehensive evaluation benchmark for multimodal large language models}.

\bibitem[{GLM et~al.(2024)GLM, Zeng, Xu, Wang, Zhang, Yin, Rojas, Feng, Zhao, Lai, Yu, Wang, Sun, Zhang, Cheng, Gui, Tang, Zhang, Li, Zhao, Wu, Zhong, Liu, Huang, Zhang, Zheng, Lu, Duan, Zhang, Cao, Yang, Tam, Zhao, Liu, Xia, Zhang, Gu, Lv, Liu, Liu, Yang, Song, Zhang, An, Xu, Niu, Yang, Li, Bai, Dong, Qi, Wang, Yang, Du, Hou, and Wang}]{glm2024chatglm}
Team GLM, Aohan Zeng, Bin Xu, Bowen Wang, Chenhui Zhang, Da~Yin, Diego Rojas, Guanyu Feng, Hanlin Zhao, Hanyu Lai, Hao Yu, Hongning Wang, Jiadai Sun, Jiajie Zhang, Jiale Cheng, Jiayi Gui, Jie Tang, Jing Zhang, Juanzi Li, Lei Zhao, Lindong Wu, Lucen Zhong, Mingdao Liu, Minlie Huang, Peng Zhang, Qinkai Zheng, Rui Lu, Shuaiqi Duan, Shudan Zhang, Shulin Cao, Shuxun Yang, Weng~Lam Tam, Wenyi Zhao, Xiao Liu, Xiao Xia, Xiaohan Zhang, Xiaotao Gu, Xin Lv, Xinghan Liu, Xinyi Liu, Xinyue Yang, Xixuan Song, Xunkai Zhang, Yifan An, Yifan Xu, Yilin Niu, Yuantao Yang, Yueyan Li, Yushi Bai, Yuxiao Dong, Zehan Qi, Zhaoyu Wang, Zhen Yang, Zhengxiao Du, Zhenyu Hou, and Zihan Wang. 2024.
\newblock \href {http://arxiv.org/abs/2406.12793} {Chatglm: A family of large language models from glm-130b to glm-4 all tools}.

\bibitem[{Grattafiori et~al.(2024)Grattafiori, Dubey, Jauhri, Pandey, Kadian, Al-Dahle, Letman, Mathur, Schelten, Vaughan et~al.}]{Dubey2024Llama31}
Aaron Grattafiori, Abhimanyu Dubey, Abhinav Jauhri, Abhinav Pandey, Abhishek Kadian, Ahmad Al-Dahle, Aiesha Letman, Akhil Mathur, Alan Schelten, Alex Vaughan, et~al. 2024.
\newblock The llama 3 herd of models.
\newblock \emph{arXiv preprint arXiv:2407.21783}.

\bibitem[{Guo and Wan(2024)}]{Guo2024PerformanceEO}
Yuhang Guo and Zhiyu Wan. 2024.
\newblock \href {https://api.semanticscholar.org/CorpusID:271940309} {Performance evaluation of multimodal large language models (llava and gpt-4-based chatgpt) in medical image classification tasks}.
\newblock \emph{2024 IEEE 12th International Conference on Healthcare Informatics (ICHI)}, pages 541--543.

\bibitem[{Huang et~al.(2024{\natexlab{a}})Huang, Yan, Li, and Peng}]{Huang2024FromLL}
Dawei Huang, Chuan Yan, Qing Li, and Xiaojiang Peng. 2024{\natexlab{a}}.
\newblock \href {https://api.semanticscholar.org/CorpusID:270409635} {From large language models to large multimodal models: A literature review}.
\newblock \emph{Applied Sciences}.

\bibitem[{Huang et~al.(2024{\natexlab{b}})Huang, Yuan, Sheng, Yang, Wu, Chen, Yang, Li, and Lin}]{Huang2024AesBench}
Yipo Huang, Quan Yuan, Xiangfei Sheng, Zhichao Yang, Haoning Wu, Pengfei Chen, Yuzhe Yang, Leida Li, and Weisi Lin. 2024{\natexlab{b}}.
\newblock Aesbench: An expert benchmark for multimodal large language models on image aesthetics perception.
\newblock \emph{arXiv preprint arXiv:2401.08276}.

\bibitem[{Hurst et~al.(2024)Hurst, Lerer, Goucher, Perelman, Ramesh, Clark, Ostrow, Welihinda, Hayes, Radford et~al.}]{hurst2024gpt}
Aaron Hurst, Adam Lerer, Adam~P Goucher, Adam Perelman, Aditya Ramesh, Aidan Clark, AJ~Ostrow, Akila Welihinda, Alan Hayes, Alec Radford, et~al. 2024.
\newblock Gpt-4o system card.
\newblock \emph{arXiv preprint arXiv:2410.21276}.

\bibitem[{Kirillov et~al.(2023)Kirillov, Mintun, Ravi, Mao, Rolland, Gustafson, Xiao, Whitehead, Berg, Lo, Doll{\'a}r, and Girshick}]{Kirillov2023SegmentA}
Alexander Kirillov, Eric Mintun, Nikhila Ravi, Hanzi Mao, Chloe Rolland, Laura Gustafson, Tete Xiao, Spencer Whitehead, Alexander~C. Berg, Wan-Yen Lo, Piotr Doll{\'a}r, and Ross~B. Girshick. 2023.
\newblock \href {https://api.semanticscholar.org/CorpusID:257952310} {Segment anything}.
\newblock \emph{2023 IEEE/CVF International Conference on Computer Vision (ICCV)}, pages 3992--4003.

\bibitem[{Ko et~al.(2022)Ko, Park, Jeon, Jo, Kim, and Seo}]{Ko2022LargescaleTG}
Hyung-Kwon Ko, Gwanmo Park, Hyeon Jeon, Jaemin Jo, Juho Kim, and Jinwook Seo. 2022.
\newblock \href {https://api.semanticscholar.org/CorpusID:252918751} {Large-scale text-to-image generation models for visual artists’ creative works}.
\newblock \emph{Proceedings of the 28th International Conference on Intelligent User Interfaces}.

\bibitem[{Li et~al.(2024)Li, Zhang, Guo, Zhang, Li, Zhang, Zhang, Zhang, Li, Liu et~al.}]{li2024llava}
Bo~Li, Yuanhan Zhang, Dong Guo, Renrui Zhang, Feng Li, Hao Zhang, Kaichen Zhang, Peiyuan Zhang, Yanwei Li, Ziwei Liu, et~al. 2024.
\newblock Llava-onevision: Easy visual task transfer.
\newblock \emph{arXiv preprint arXiv:2408.03326}.

\bibitem[{Li et~al.(2023)Li, Ge, Ge, Wang, Wang, Zhang, and Shan}]{Li2023SEEDBench2}
Bohao Li, Yuying Ge, Yixiao Ge, Guangzhi Wang, Rui Wang, Ruimao Zhang, and Ying Shan. 2023.
\newblock \href {https://api.semanticscholar.org/CorpusID:265498814} {Seed-bench-2: Benchmarking multimodal large language models}.
\newblock \emph{ArXiv}, abs/2311.17092.

\bibitem[{Liu et~al.(2024)Liu, Li, Li, Li, Zhang, Shen, and Lee}]{liu2024llavanext}
Haotian Liu, Chunyuan Li, Yuheng Li, Bo~Li, Yuanhan Zhang, Sheng Shen, and Yong~Jae Lee. 2024.
\newblock \href {https://llava-vl.github.io/blog/2024-01-30-llava-next/} {Llava-next: Improved reasoning, ocr, and world knowledge}.

\bibitem[{Lu et~al.(2024{\natexlab{a}})Lu, Bansal, Xia, Liu, Li, Hajishirzi, Cheng, Chang, Galley, and Gao}]{lu2024mathvista}
Pan Lu, Hritik Bansal, Tony Xia, Jiacheng Liu, Chunyuan Li, Hannaneh Hajishirzi, Hao Cheng, Kai-Wei Chang, Michel Galley, and Jianfeng Gao. 2024{\natexlab{a}}.
\newblock Mathvista: Evaluating mathematical reasoning of foundation models in visual contexts.
\newblock In \emph{International Conference on Learning Representations (ICLR)}.

\bibitem[{Lu et~al.(2024{\natexlab{b}})Lu, Li, Chen, Xu, Luo, Zhang, and Ye}]{lu2024ovis}
Shiyin Lu, Yang Li, Qing-Guo Chen, Zhao Xu, Weihua Luo, Kaifu Zhang, and Han-Jia Ye. 2024{\natexlab{b}}.
\newblock Ovis: Structural embedding alignment for multimodal large language model.
\newblock \emph{arXiv:2405.20797}.

\bibitem[{Masry et~al.(2022)Masry, Long, Tan, Joty, and Hoque}]{Masry2022ChartQA}
Ahmed Masry, Do~Xuan Long, Jia~Qing Tan, Shafiq~R. Joty, and Enamul Hoque. 2022.
\newblock \href {https://api.semanticscholar.org/CorpusID:247593713} {Chartqa: A benchmark for question answering about charts with visual and logical reasoning}.
\newblock \emph{ArXiv}, abs/2203.10244.

\bibitem[{Qiao et~al.(2024)Qiao, Tan, Dong, Wu, Sun, Song, GongQue, Lei, Wei, Zhang et~al.}]{qiao2024we}
Runqi Qiao, Qiuna Tan, Guanting Dong, Minhui Wu, Chong Sun, Xiaoshuai Song, Zhuoma GongQue, Shanglin Lei, Zhe Wei, Miaoxuan Zhang, et~al. 2024.
\newblock We-math: Does your large multimodal model achieve human-like mathematical reasoning?
\newblock \emph{arXiv preprint arXiv:2407.01284}.

\bibitem[{Ramesh et~al.(2021)Ramesh, Pavlov, Goh, Gray, Voss, Radford, Chen, and Sutskever}]{Ramesh2021ZeroShotTG}
Aditya Ramesh, Mikhail Pavlov, Gabriel Goh, Scott Gray, Chelsea Voss, Alec Radford, Mark Chen, and Ilya Sutskever. 2021.
\newblock \href {https://api.semanticscholar.org/CorpusID:232035663} {Zero-shot text-to-image generation}.
\newblock \emph{ArXiv}, abs/2102.12092.

\bibitem[{Sepehri et~al.(2024)Sepehri, Fabian, Soltanolkotabi, and Soltanolkotabi}]{sepehri2024mediconfusion}
Mohammad~Shahab Sepehri, Zalan Fabian, Maryam Soltanolkotabi, and Mahdi Soltanolkotabi. 2024.
\newblock Mediconfusion: Can you trust your ai radiologist? probing the reliability of multimodal medical foundation models.
\newblock \emph{arXiv preprint arXiv:2409.15477}.

\bibitem[{Shen et~al.(2023)Shen, Fu, Chen, Zhang, Li, Sun, Wu, Lin, and Ji}]{Shen2023AligningAP}
Yunhang Shen, Chaoyou Fu, Peixian Chen, Mengdan Zhang, Ke~Li, Xing Sun, Yunsheng Wu, Shaohui Lin, and Rongrong Ji. 2023.
\newblock \href {https://api.semanticscholar.org/CorpusID:265609653} {Aligning and prompting everything all at once for universal visual perception}.
\newblock \emph{ArXiv}, abs/2312.02153.

\bibitem[{Song et~al.(2024)Song, Chen, Chen, Yu, Wan, and Wang}]{song2024milebench}
Dingjie Song, Shunian Chen, Guiming~Hardy Chen, Fei Yu, Xiang Wan, and Benyou Wang. 2024.
\newblock Milebench: Benchmarking mllms in long context.
\newblock \emph{arXiv preprint arXiv:2404.18532}.

\bibitem[{Team(2025)}]{gemma_2025}
Gemma Team. 2025.
\newblock \href {https://goo.gle/Gemma3Report} {Gemma 3}.

\bibitem[{Team et~al.(2025)Team, Du, Yin, Xing, Qu, Wang, Chen, Zhang, Du, Wei, Wang, Zhang, Du, Wang, Yuan, Lu, Li, Sung, Wei, Lai, Zhu, Ding, Hu, Yang, Zhang, Wu, Yao, Lu, Wang, Gao, Zheng, Li, Su, Wang, Deng, Qiu, Xie, Wang, Liu, Yan, Ouyang, Chen, Sui, Yu, Dong, Dong, Xu, Cheng, Gu, Zhou, Liu, Cao, Yu, Song, Bai, Song, He, Huang, Xu et~al.}]{kimiteam2025kimivl}
Kimi Team, Angang Du, Bohong Yin, Bowei Xing, Bowen Qu, Bowen Wang, Cheng Chen, Chenlin Zhang, Chenzhuang Du, Chu Wei, Congcong Wang, Dehao Zhang, Dikang Du, Dongliang Wang, Enming Yuan, Enzhe Lu, Fang Li, Flood Sung, Guangda Wei, Guokun Lai, Han Zhu, Hao Ding, Hao Hu, Hao Yang, Hao Zhang, Haoning Wu, Haotian Yao, Haoyu Lu, Heng Wang, Hongcheng Gao, Huabin Zheng, Jiaming Li, Jianlin Su, Jianzhou Wang, Jiaqi Deng, Jiezhong Qiu, Jin Xie, Jinhong Wang, Jingyuan Liu, Junjie Yan, Kun Ouyang, Liang Chen, Lin Sui, Longhui Yu, Mengfan Dong, Mengnan Dong, Nuo Xu, Pengyu Cheng, Qizheng Gu, Runjie Zhou, Shaowei Liu, Sihan Cao, Tao Yu, Tianhui Song, Tongtong Bai, Wei Song, Weiran He, Weixiao Huang, Weixin Xu, et~al. 2025.
\newblock Kimi-vl technical report.
\newblock \emph{arXiv preprint arXiv:2504.07491}.

\bibitem[{Team(2024)}]{qwen2.5}
Qwen Team. 2024.
\newblock \href {https://qwenlm.github.io/blog/qwen2.5/} {Qwen2.5: A party of foundation models}.

\bibitem[{Tong et~al.(2024)Tong, Brown, Wu, Woo, Middepogu, Akula, Yang, Yang, Iyer, Pan, Wang, Fergus, LeCun, and Xie}]{tong2024cambrian1}
Shengbang Tong, Ellis Brown, Penghao Wu, Sanghyun Woo, Manoj Middepogu, Sai~Charitha Akula, Jihan Yang, Shusheng Yang, Adithya Iyer, Xichen Pan, Austin Wang, Rob Fergus, Yann LeCun, and Saining Xie. 2024.
\newblock \href {http://arxiv.org/abs/2406.16860} {Cambrian-1: A fully open, vision-centric exploration of multimodal llms}.

\bibitem[{Wang et~al.(2023)Wang, Lv, Yu, Hong, Qi, Wang, Ji, Yang, Zhao, Song, Xu, Xu, Li, Dong, Ding, and Tang}]{wang2023cogvlm}
Weihan Wang, Qingsong Lv, Wenmeng Yu, Wenyi Hong, Ji~Qi, Yan Wang, Junhui Ji, Zhuoyi Yang, Lei Zhao, Xixuan Song, Jiazheng Xu, Bin Xu, Juanzi Li, Yuxiao Dong, Ming Ding, and Jie Tang. 2023.
\newblock \href {http://arxiv.org/abs/2311.03079} {Cogvlm: Visual expert for pretrained language models}.

\bibitem[{Wang et~al.(2024)Wang, Xia, He, Chen, Liu, Zhu, Liang, Wu, Liu, Malladi, Chevalier, Arora, and Chen}]{Wang2024CharXiv}
Zirui Wang, Mengzhou Xia, Luxi He, Howard Chen, Yitao Liu, Richard Zhu, Kaiqu Liang, Xindi Wu, Haotian Liu, Sadhika Malladi, Alexis Chevalier, Sanjeev Arora, and Danqi Chen. 2024.
\newblock \href {https://api.semanticscholar.org/CorpusID:270737638} {Charxiv: Charting gaps in realistic chart understanding in multimodal llms}.
\newblock \emph{ArXiv}, abs/2406.18521.

\bibitem[{Wu et~al.(2024)Wu, Chen, Pan, Liu, Liu, Dai, Gao, Ma, Wu, Wang, Xie, Wu, Hu, Wang, Sun, Li, Piao, Guan, Liu, Xie, You, Dong, Yu, Zhang, Zhao, Wang, and Ruan}]{wu2024deepseekvl2}
Zhiyu Wu, Xiaokang Chen, Zizheng Pan, Xingchao Liu, Wen Liu, Damai Dai, Huazuo Gao, Yiyang Ma, Chengyue Wu, Bingxuan Wang, Zhenda Xie, Yu~Wu, Kai Hu, Jiawei Wang, Yaofeng Sun, Yukun Li, Yishi Piao, Kang Guan, Aixin Liu, Xin Xie, Yuxiang You, Kai Dong, Xingkai Yu, Haowei Zhang, Liang Zhao, Yisong Wang, and Chong Ruan. 2024.
\newblock \href {http://arxiv.org/abs/2412.10302} {Deepseek-vl2: Mixture-of-experts vision-language models for advanced multimodal understanding}.

\bibitem[{Xia et~al.(2024)Xia, Zhang, Ye, Yan, Liu, Zhou, Chen, Ye, Dou, Shi et~al.}]{xia2024chartx}
Renqiu Xia, Bo~Zhang, Hancheng Ye, Xiangchao Yan, Qi~Liu, Hongbin Zhou, Zijun Chen, Peng Ye, Min Dou, Botian Shi, et~al. 2024.
\newblock Chartx \& chartvlm: A versatile benchmark and foundation model for complicated chart reasoning.
\newblock \emph{arXiv preprint arXiv:2402.12185}.

\bibitem[{Yao et~al.(2024)Yao, Yu, Zhang, Wang, Cui, Zhu, Cai, Li, Zhao, He et~al.}]{yao2024minicpm}
Yuan Yao, Tianyu Yu, Ao~Zhang, Chongyi Wang, Junbo Cui, Hongji Zhu, Tianchi Cai, Haoyu Li, Weilin Zhao, Zhihui He, et~al. 2024.
\newblock Minicpm-v: A gpt-4v level mllm on your phone.
\newblock \emph{arXiv preprint arXiv:2408.01800}.

\bibitem[{Ying et~al.(2024)Ying, Meng, Wang, Li, Lin, Yang, Zhang, Zhang, Lin, Liu, Lei, Lu, Chen, Xu, Zhang, Zhang, Gao, Wang, Qiao, Luo, Zhang, and Shao}]{mmtbench}
Kaining Ying, Fanqing Meng, Jin Wang, Zhiqian Li, Han Lin, Yue Yang, Hao Zhang, Wenbo Zhang, Yuqi Lin, Shuo Liu, Jiayi Lei, Quanfeng Lu, Runjian Chen, Peng Xu, Renrui Zhang, Haozhe Zhang, Peng Gao, Yali Wang, Yu~Qiao, Ping Luo, Kaipeng Zhang, and Wenqi Shao. 2024.
\newblock \href {http://arxiv.org/abs/2404.16006} {Mmt-bench: A comprehensive multimodal benchmark for evaluating large vision-language models towards multitask agi}.

\bibitem[{Young et~al.(2024)Young, Chen, Li, Huang, Zhang, Zhang, Wang, Li, Zhu, Chen et~al.}]{young2024yi}
Alex Young, Bei Chen, Chao Li, Chengen Huang, Ge~Zhang, Guanwei Zhang, Guoyin Wang, Heng Li, Jiangcheng Zhu, Jianqun Chen, et~al. 2024.
\newblock Yi: Open foundation models by 01. ai.
\newblock \emph{arXiv preprint arXiv:2403.04652}.

\bibitem[{Yu et~al.(2024)Yu, Yang, Li, Wang, Lin, Liu, Wang, and Wang}]{yu2024mm}
Weihao Yu, Zhengyuan Yang, Linjie Li, Jianfeng Wang, Kevin Lin, Zicheng Liu, Xinchao Wang, and Lijuan Wang. 2024.
\newblock Mm-vet: Evaluating large multimodal models for integrated capabilities.
\newblock In \emph{International conference on machine learning}. PMLR.

\bibitem[{Yue et~al.(2024)Yue, Zheng, Ni, Wang, Zhang, Tong, Sun, Yu, Zhang, Sun, Su, Chen, and Neubig}]{yue2024mmmu}
Xiang Yue, Tianyu Zheng, Yuansheng Ni, Yubo Wang, Kai Zhang, Shengbang Tong, Yuxuan Sun, Botao Yu, Ge~Zhang, Huan Sun, Yu~Su, Wenhu Chen, and Graham Neubig. 2024.
\newblock Mmmu-pro: A more robust multi-discipline multimodal understanding benchmark.
\newblock \emph{arXiv preprint arXiv:2409.02813}.

\bibitem[{Zeng et~al.(2022)Zeng, Liu, Du, Wang, Lai, Ding, Yang, Xu, Zheng, Xia, Tam, Ma, Xue, Zhai, Chen, Zhang, Dong, and Tang}]{Zeng2022GLM130BAO}
Aohan Zeng, Xiao Liu, Zhengxiao Du, Zihan Wang, Hanyu Lai, Ming Ding, Zhuoyi Yang, Yifan Xu, Wendi Zheng, Xiao Xia, Weng~Lam Tam, Zixuan Ma, Yufei Xue, Jidong Zhai, Wenguang Chen, P.~Zhang, Yuxiao Dong, and Jie Tang. 2022.
\newblock \href {https://api.semanticscholar.org/CorpusID:252715691} {Glm-130b: An open bilingual pre-trained model}.
\newblock \emph{ArXiv}, abs/2210.02414.

\bibitem[{Zhai et~al.(2023)Zhai, Mustafa, Kolesnikov, and Beyer}]{zhai2023sigmoid}
Xiaohua Zhai, Basil Mustafa, Alexander Kolesnikov, and Lucas Beyer. 2023.
\newblock \href {http://arxiv.org/abs/2303.15343} {Sigmoid loss for language image pre-training}.

\bibitem[{Zhang et~al.(2022)Zhang, Li, Liu, Zhang, Su, Zhu, shuan Ni, and yeung Shum}]{Zhang2022DINODW}
Hao Zhang, Feng Li, Shilong Liu, Lei Zhang, Hang Su, Jun-Juan Zhu, Lionel~Ming shuan Ni, and Heung yeung Shum. 2022.
\newblock \href {https://api.semanticscholar.org/CorpusID:247292561} {Dino: Detr with improved denoising anchor boxes for end-to-end object detection}.
\newblock \emph{ArXiv}, abs/2203.03605.

\bibitem[{Zhang et~al.(2024)Zhang, Jiang, Zhang, Lin, Guo, Qiu, Zhou, Lu, Chang, Qiao et~al.}]{zhang2024mathverse}
Renrui Zhang, Dongzhi Jiang, Yichi Zhang, Haokun Lin, Ziyu Guo, Pengshuo Qiu, Aojun Zhou, Pan Lu, Kai-Wei Chang, Yu~Qiao, et~al. 2024.
\newblock Mathverse: Does your multi-modal llm truly see the diagrams in visual math problems?
\newblock In \emph{European Conference on Computer Vision}, pages 169--186. Springer.

\bibitem[{Zhu et~al.(2025)Zhu, Wang, Chen, Liu, Ye, Gu, Duan, Tian, Su, Shao et~al.}]{zhu2025internvl3}
Jinguo Zhu, Weiyun Wang, Zhe Chen, Zhaoyang Liu, Shenglong Ye, Lixin Gu, Yuchen Duan, Hao Tian, Weijie Su, Jie Shao, et~al. 2025.
\newblock Internvl3: Exploring advanced training and test-time recipes for open-source multimodal models.
\newblock \emph{arXiv preprint arXiv:2504.10479}.

\bibitem[{Zuo et~al.(2025)Zuo, Qu, Li, Chen, Zhu, Hua, Zhang, Ding, and Zhou}]{zuo2025medxpertqa}
Yuxin Zuo, Shang Qu, Yifei Li, Zhangren Chen, Xuekai Zhu, Ermo Hua, Kaiyan Zhang, Ning Ding, and Bowen Zhou. 2025.
\newblock Medxpertqa: Benchmarking expert-level medical reasoning and understanding.
\newblock \emph{arXiv preprint arXiv:2501.18362}.

\end{thebibliography}

\appendix
\section{Prompts}\label{appendixA}
\subsection{Prompt for Rewriting Descriptions}\label{appendixA:prompt1}
The following Promt is used in LLaMA3.3-70B-Instruct model to rewrite the context text description of the image.
\begin{questionbox}[grey]
\small
\textbf{SYSTEM PROMPT:} \\
Act like an expert with extensive experience writing in the field of astrophysics. \\

\textbf{Objective:} \\
You will be provided with the \texttt{[CAPTION]}, and \texttt{[CONTEXT]} of an image mentioned in the paper. Meanwhile, the \texttt{[TITLE]} and \texttt{[ABSTRACT]} of the paper are also provided as background information to you. Your task is to generate a concise, precise, and scholarly description of the image, reflecting its content and relevance within the scientific discourse of the paper. Your answer will serve senior scholars, please describe it in a formal and scholarly manner.
\\

\textbf{Detailed Instructions:}
\begin{enumerate}
    \item \textbf{Content Analysis:}
    \begin{itemize}
        \item Carefully review \texttt{[CAPTION]}, and \texttt{[CONTEXT]} of the image, determining target image and ensuring a thorough understanding of its significance and details. 
        \item Examine the \texttt{[TITLE]} and \texttt{[ABSTRACT]} of the paper and use those background informations if necessary.
    \end{itemize}
    
    \item \textbf{Formatting and Content Restrictions:}
    \begin{itemize}
        \item Ensure all LaTeX formats are deleted except for mathematical formulas.
        \item Ignore any content related to unknown objects in the paper, such as other formulas, images or sections, and do not summarize them.
        \item If the content you are given is not related to the target image, ignore it and do not summarize it.
        \item When you refer to the target image, use expressions such as "The image" or "The figure", instead of "Figure~\texttt{\textbackslash ref\{?\}}" or "Figure ?".
    \end{itemize}
    
    \item \textbf{Writing the Description:}
    \begin{itemize}
        \item Formulate a comprehensive and scholarly description of the image using the gathered information.
    \end{itemize}
\end{enumerate}

\textbf{Output Format:} \\
\{\\
  "description": "The description you generated here"\\
\}

\vspace{0.5cm}
\textbf{USER PROMPT:} \\
Please give your description based on the following informations: \\
\texttt{[CAPTION]}: \{caption\} \\
\texttt{[CONTEXT]}: \{context\} \\
\texttt{[TITLE]}: \{title\} \\
\texttt{[ABSTRACT]}: \{abstract\}
\end{questionbox}
\newpage
\subsection{Prompt for Question Generation}\label{appendixA:prompt2}
The following Promt is used in the InternVL2.5-78B model to generate high-quality multi-modal multiple-choice questions in the field of astronomy.
\begin{questionbox}[boxcolor1]
\small
Act like a domain expert in astronomy education, with extensive experience in designing high-level exam questions that assess advanced conceptual and analytical skills. \\

\textbf{Objective:}\\
You will receive an image and its associated descriptions. Your task is to generate a multiple-choice question (include one correct option and three plausible but incorrect options) at a professional level that tests the respondent's ability to analyze images and apply comprehensive astronomical knowledge.  \\

\textbf{Detailed Instructions:}
\begin{enumerate}
    \item \textbf{Image and Description Analysis:}
    \begin{itemize}
        \item View the \texttt{[IMAGE]} provided thoroughly, noting any important subjects, features, and text, etc.
        \item Read the \texttt{[IMAGE DESCRIPTIONS]} carefully to determine the relationship between the description and the image, and consider the astronomical knowledge involved.
    \end{itemize}

    \item \textbf{Question Design:}
    \begin{itemize}
        \item Create a question that requires image analysis, astronomical knowledge, and in-depth analysis to solve, ensuring it does not provide hints.
    \end{itemize}
    
    \item \textbf{Create Answer Choices:}
    \begin{itemize}
        \item Determine an answer to the question as the correct option.
        \item Develop three plausible but incorrect options.
    \end{itemize}
    
    \item \textbf{Explanation of the Correct Answer:}
    \begin{itemize}
        \item Provide a detailed explanation for why the correct answer is accurate.
        \item Optionally, briefly state why each incorrect option is misleading or incorrect.

    \end{itemize}
\end{enumerate}

\textbf{Output Format:} \\
\{\\
  "question": "Your image-based astronomical question here",\\
  "options": \{\\
    "A": "Option A content",\\
    "B": "Option B content",\\
    "C": "Option C content",\\
    "D": "Option D content"\\
  \},\\
  "answer": "Correct option letter" \\
  "explanation": "Brief justification for the correct answer."
\}\\

\textbf{Input:} \\ 
Please generate the question based on the following: \\
\texttt{[IMAGE DESCRIPTIONS]:\{image\_descriptions\}}
\end{questionbox}

\section{Model Evaluation Examples}\label{appendixB}
This section presents 18 example questions on random sampling across varying levels of difficulty, with three questions selected from each subdomain. 

\subsection{Solar and Stellar Astrophysics (SR)}\label{appendixB:SR}

\begin{examplebox}[White]{\text{Correct responses: 24/25 models}}
\small
\begin{center}
    \includegraphics[width=\textwidth]{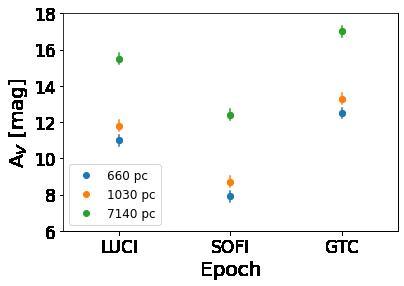}
\end{center}

\textbf{Question:} Which epoch shows the highest visual extinction ($A_V$) for Gaial8cjb at distance of 7140 pc?

\textbf{Option:}
\begin{enumerate}[label=(\Alph*)]
    \item LUCI
    \item SOFI
    \item GTC
    \item None of the above
\end{enumerate}
\textbf{Answer:} \textcolor{ForestGreen}{C}
\vspace{0.5em}

\textbf{Ovis2-34B:} \textcolor{ForestGreen}{C}
\vspace{0.5em}

\textbf{ChatGPT-4o: } \textcolor{ForestGreen}{C}
\vspace{0.5em}

\textbf{Doubao-1.5-vision-pro: } \textcolor{ForestGreen}{C}
\vspace{0.5em}

\textbf{InternVL3-38B: } \textcolor{ForestGreen}{C}
\vspace{0.5em}

\textbf{Qwen2.5-VL-72B: } \textcolor{ForestGreen}{C}
\vspace{0.5em}

\textbf{LLaVA\_Onevision\_72B: } \textcolor{red}{B}
\vspace{0.5em}

\textbf{Gemma3-12B: } \textcolor{ForestGreen}{C}
\end{examplebox}
\begin{center}
    Figure B1: Case 1 of AstroMMBench in SR subdomain.
\end{center}

\begin{examplebox}[White]{\text{Correct responses: 17/25 models}}
\small
\begin{center}
    \includegraphics[width=\textwidth]{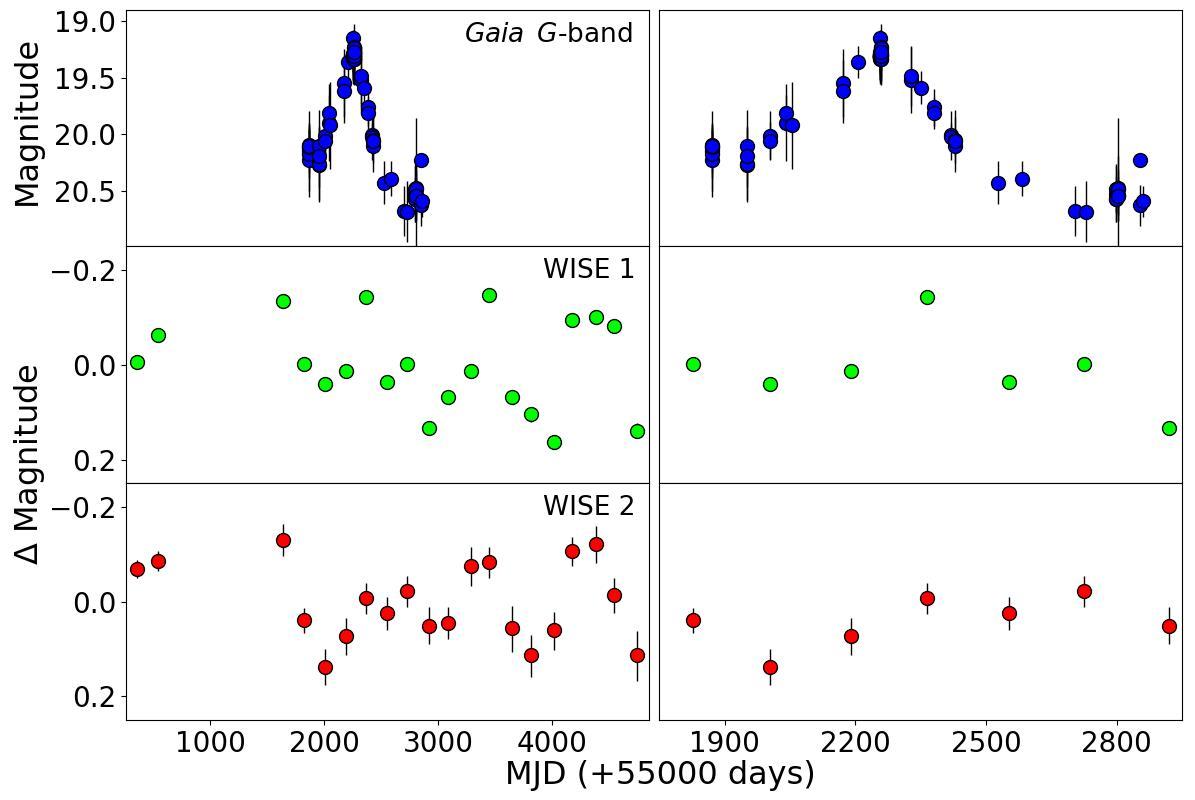}
\end{center}

\textbf{Question:} Which of the following best describes the periodicity of the light curve for NGC300-59 in the Gaia G-band?

\textbf{Option:}
\begin{enumerate}[label=(\Alph*)]
    \item Approximately 500 days
    \item Approximately 1000 days
    \item Approximately 2000 days
    \item Approximately 4000 days
\end{enumerate}
\textbf{Answer:} \textcolor{ForestGreen}{B}
\vspace{0.5em}

\textbf{Ovis2-34B:} \textcolor{red}{C}
\vspace{0.5em}

\textbf{ChatGPT-4o: } \textcolor{ForestGreen}{B}
\vspace{0.5em}

\textbf{Doubao-1.5-vision-pro: } \textcolor{red}{C}
\vspace{0.5em}

\textbf{InternVL3-38B: } \textcolor{ForestGreen}{B}
\vspace{0.5em}

\textbf{Qwen2.5-VL-72B: } \textcolor{red}{C}
\vspace{0.5em}

\textbf{LLaVA\_Onevision\_72B: } \textcolor{ForestGreen}{B}
\vspace{0.5em}

\textbf{Gemma3-12B: } \textcolor{ForestGreen}{B}

\end{examplebox}
\begin{center}
Figure B2: Case 2 of AstroMMBench in SR subdomain.
\end{center}

\begin{examplebox}[White]{\text{Correct responses: 4/25 models}}
\small
\begin{center}
    \includegraphics[width=\textwidth, height=3.5cm]{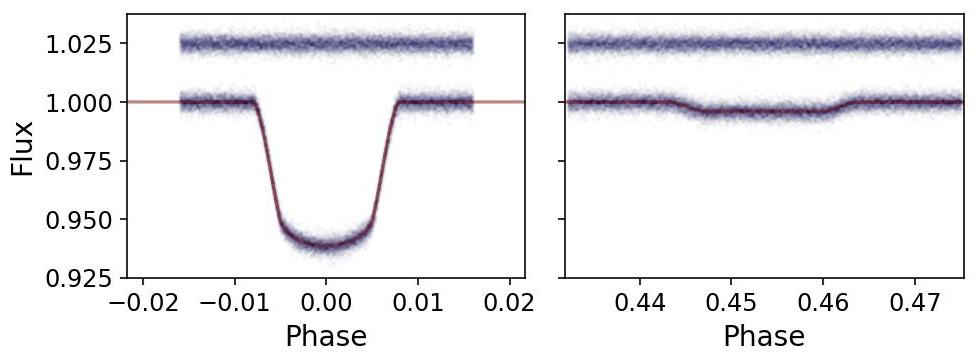}
\end{center}

\textbf{Question:} What is the primary feature observed in the light curve of the EBLM J0608-59 system?

\textbf{Option:}
\begin{enumerate}[label=(\Alph*)]
    \item A single transit event
    \item A double eclipse event
    \item A single eclipse event
    \item A continuous out-of-eclipse variation
\end{enumerate}
\textbf{Answer:} \textcolor{ForestGreen}{B}
\vspace{0.5em}

\textbf{Ovis2-34B:} \textcolor{red}{A}
\vspace{0.5em}

\textbf{ChatGPT-4o: } \textcolor{red}{C}
\vspace{0.5em}

\textbf{Doubao-1.5-vision-pro: } \textcolor{red}{C}
\vspace{0.5em}

\textbf{InternVL3-38B: } \textcolor{ForestGreen}{B}
\vspace{0.5em}

\textbf{Qwen2.5-VL-72B: } \textcolor{red}{C}
\vspace{0.5em}

\textbf{LLaVA\_Onevision\_72B: } \textcolor{red}{A}
\vspace{0.5em}

\textbf{Gemma3-12B: } \textcolor{red}{A}

\end{examplebox}
\begin{center}
Figure B3: Case 3 of AstroMMBench in SR subdomain.
\end{center}

\subsection{Instrumentation and Methods for Astrophysics (IM)}\label{appendixB:IM}
\begin{examplebox}[White]{\text{Correct responses: 24/25 models}}
\small
\begin{center}
    \includegraphics[width=\textwidth]{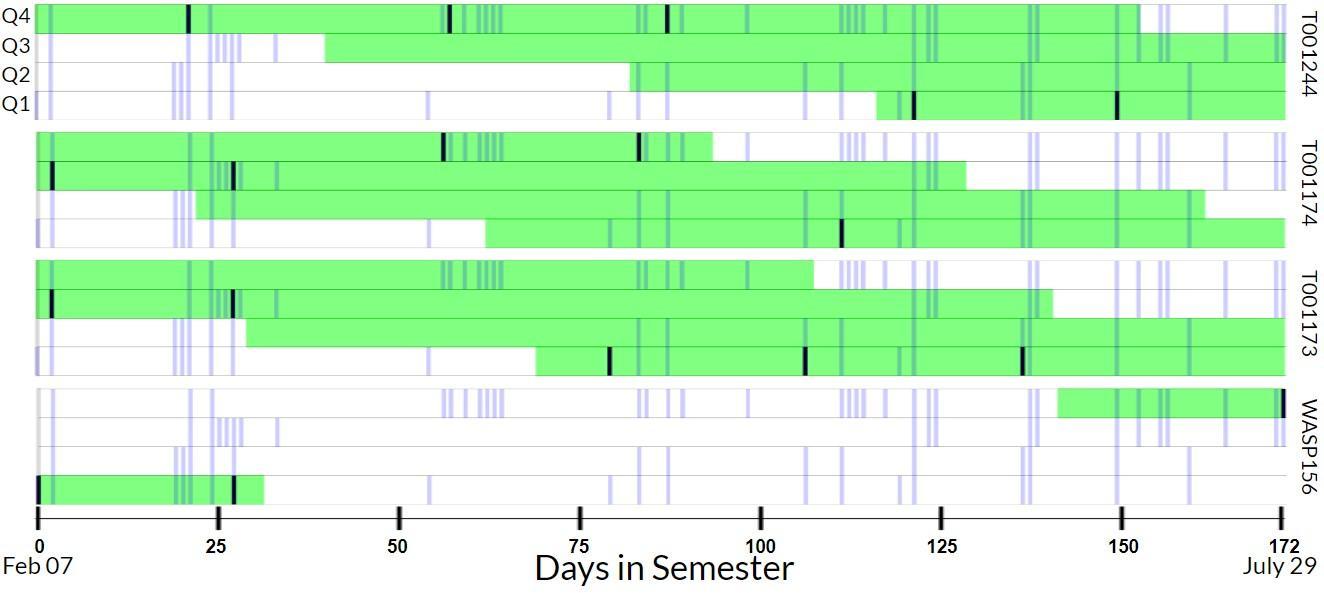}
\end{center}

\textbf{Question:} Which of the following targets has the most limited visibility in the 2023A California Planet Search(CPs)simulation?

\textbf{Option:}
\begin{enumerate}[label=(\Alph*)]
    \item Target 100244
    \item Target 100174
    \item Target 100173
    \item Target WASP159
\end{enumerate}
\textbf{Answer:} \textcolor{ForestGreen}{D}
\vspace{0.5em}

\textbf{Ovis2-34B:} \textcolor{ForestGreen}{D}
\vspace{0.5em}

\textbf{ChatGPT-4o: } \textcolor{ForestGreen}{D}
\vspace{0.5em}

\textbf{Doubao-1.5-vision-pro: } \textcolor{ForestGreen}{D}
\vspace{0.5em}

\textbf{InternVL3-38B: } \textcolor{ForestGreen}{D}
\vspace{0.5em}

\textbf{Qwen2.5-VL-72B: } \textcolor{ForestGreen}{D}
\vspace{0.5em}

\textbf{LLaVA\_Onevision\_72B: } \textcolor{ForestGreen}{D}
\vspace{0.5em}

\textbf{Gemma3-12B: } \textcolor{ForestGreen}{D}

\end{examplebox}
\begin{center}
Figure B4: Case 4 of AstroMMBench in IM subdomain.
\end{center}

\begin{examplebox}[White]{\text{Correct responses: 16/25 models}}
\small
\begin{center}
    \includegraphics[width=0.9\textwidth]{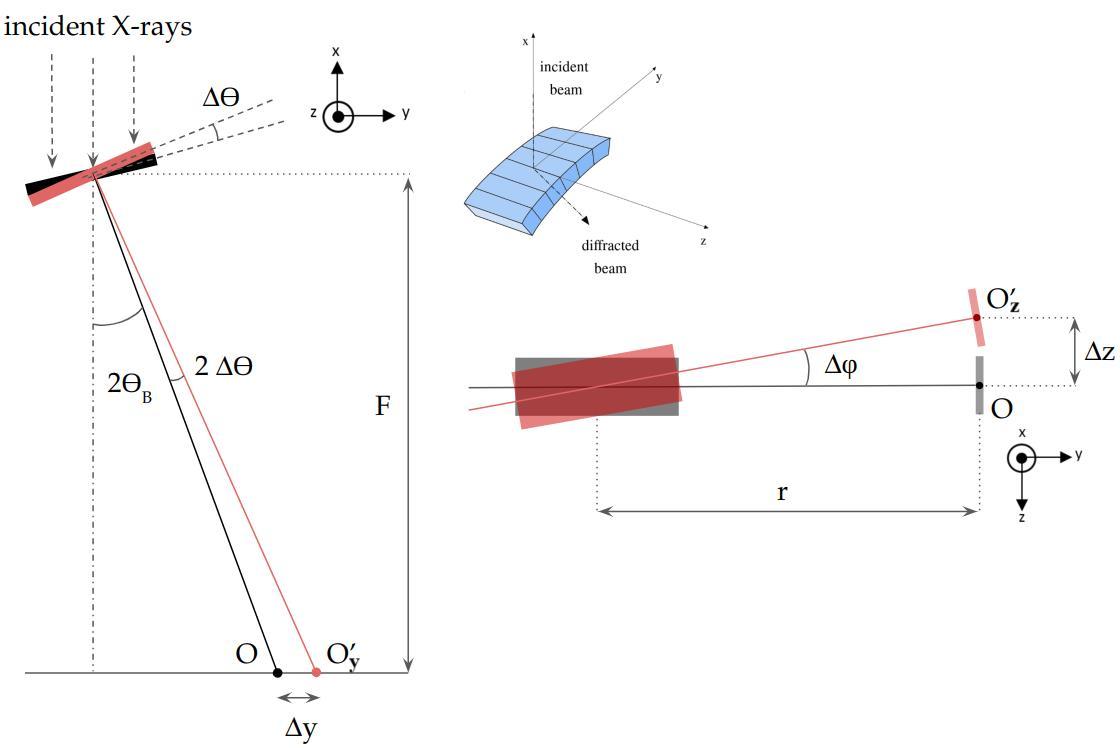}
\end{center}

\textbf{Question:} What is the primary consequence of a misalignment in the Bragg angle $ \theta (\Delta \theta) $ on the diffracted X-ray beam in the Laue lens?

\textbf{Option:}
\begin{enumerate}[label=(\Alph*)]
    \item A shift in the diffracted signal along the z-axis
    \item A shift in the diffracted signal along the y-axis
    \item A change in the intensity of the diffracted beam
    \item A change in the polarization of the diffracted beam
\end{enumerate}
\textbf{Answer:} \textcolor{ForestGreen}{B}
\vspace{0.5em}

\textbf{Ovis2-34B:} \textcolor{ForestGreen}{B}
\vspace{0.5em}

\textbf{ChatGPT-4o: } \textcolor{ForestGreen}{B}
\vspace{0.5em}

\textbf{Doubao-1.5-vision-pro: } \textcolor{ForestGreen}{B}
\vspace{0.5em}

\textbf{InternVL3-38B: } \textcolor{ForestGreen}{B}
\vspace{0.5em}

\textbf{Qwen2.5-VL-72B: } \textcolor{ForestGreen}{B}
\vspace{0.5em}

\textbf{LLaVA\_Onevision\_72B: } \textcolor{red}{A}
\vspace{0.5em}

\textbf{Gemma3-12B: } \textcolor{red}{A}

\end{examplebox}
\begin{center}
Figure B5: Case 5 of AstroMMBench in IM subdomain.
\end{center}

\begin{examplebox}[White]{\text{Correct responses: 1/25 models}}
\small
\begin{center}
    \includegraphics[width=0.9\textwidth]{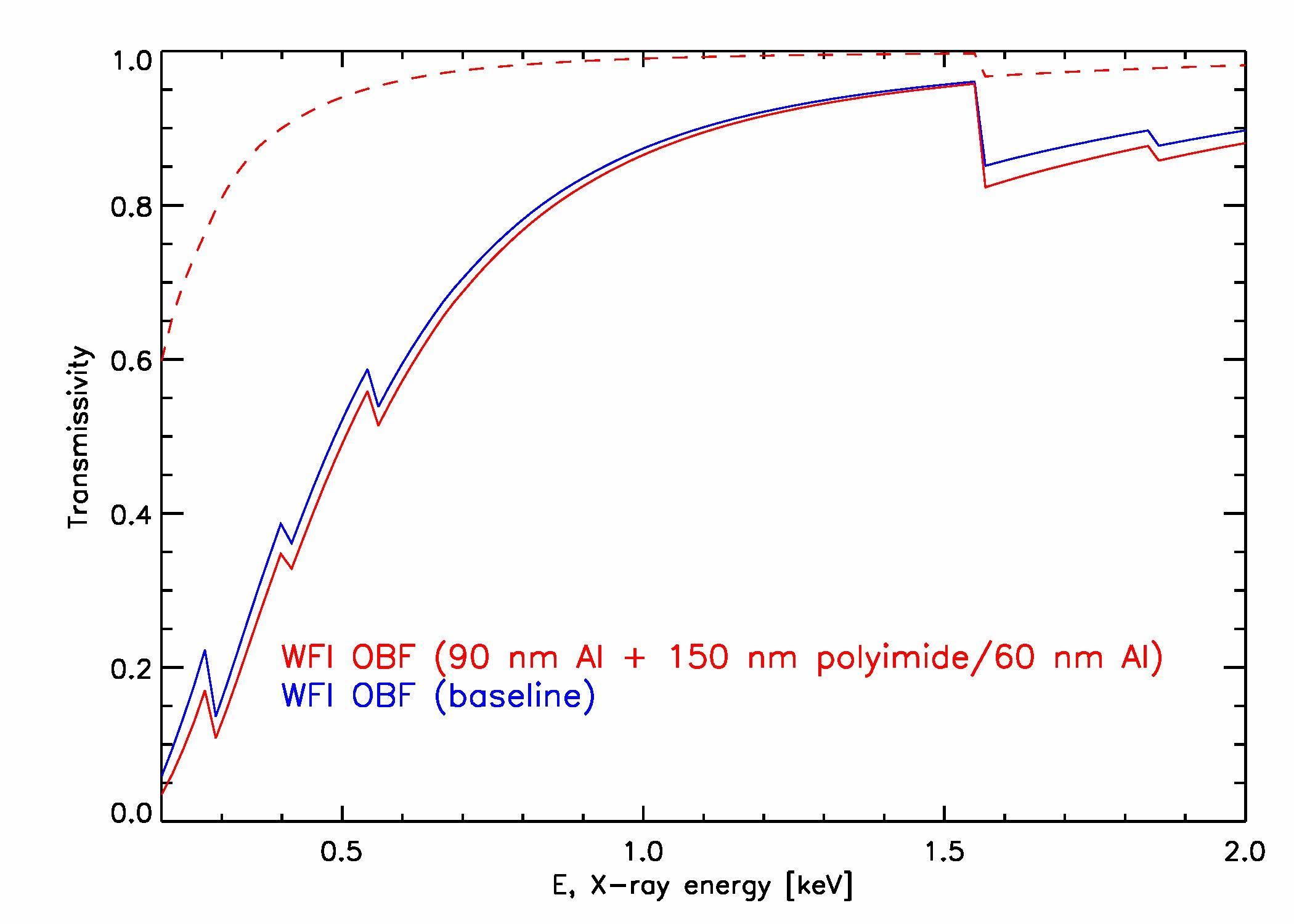}
\end{center}

\textbf{Question:} Which of the following statements is true regarding the transmissivity of the WFI OBF configurations shown in the image?

\textbf{Option:} 
\begin{enumerate}[label=(\Alph*)]
    \item The baseline WFI OBF has higher transmissivity across all energy bands compared to the 90 nm Al + 150 nm polyimide/60 nm Al configuration.
    \item The 90 nm Al + 150 nm polyimide/60 nm Al configuration has higher transmissivity below 0.5 keV compared to the baseline WFI OBF.
    \item The transmissivity ratio between the two configurations is constant across the entire energy range.
    \item The 90 nm Al + 150 nm polyimide/60 nm Al configuration has higher transmissivity above 1.5 keV compared to the baseline WFI OBF.
\end{enumerate}

\textbf{Answer:} \textcolor{ForestGreen}{A}
\vspace{0.5em}

\textbf{Ovis2-34B:} \textcolor{red}{D}
\vspace{0.5em}

\textbf{ChatGPT-4o: } \textcolor{red}{B}
\vspace{0.5em}

\textbf{Doubao-1.5-vision-pro: } \textcolor{red}{D}
\vspace{0.5em}

\textbf{InternVL3-38B: } \textcolor{red}{B}
\vspace{0.5em}

\textbf{Qwen2.5-VL-72B: } \textcolor{red}{B}
\vspace{0.5em}

\textbf{LLaVA\_Onevision\_72B: } \textcolor{red}{D}
\vspace{0.5em}

\textbf{Gemma3-12B: } \textcolor{red}{B}

\end{examplebox}
\begin{center}
Figure B6: Case 6 of AstroMMBench in IM subdomain.
\end{center}

\subsection{Earth and Planetary Astrophysics (EP)}\label{appendixB:EP}

\begin{examplebox}[White]{\text{Correct responses: 24/25 models}}
\small
\begin{center}
    \includegraphics[width=\textwidth]{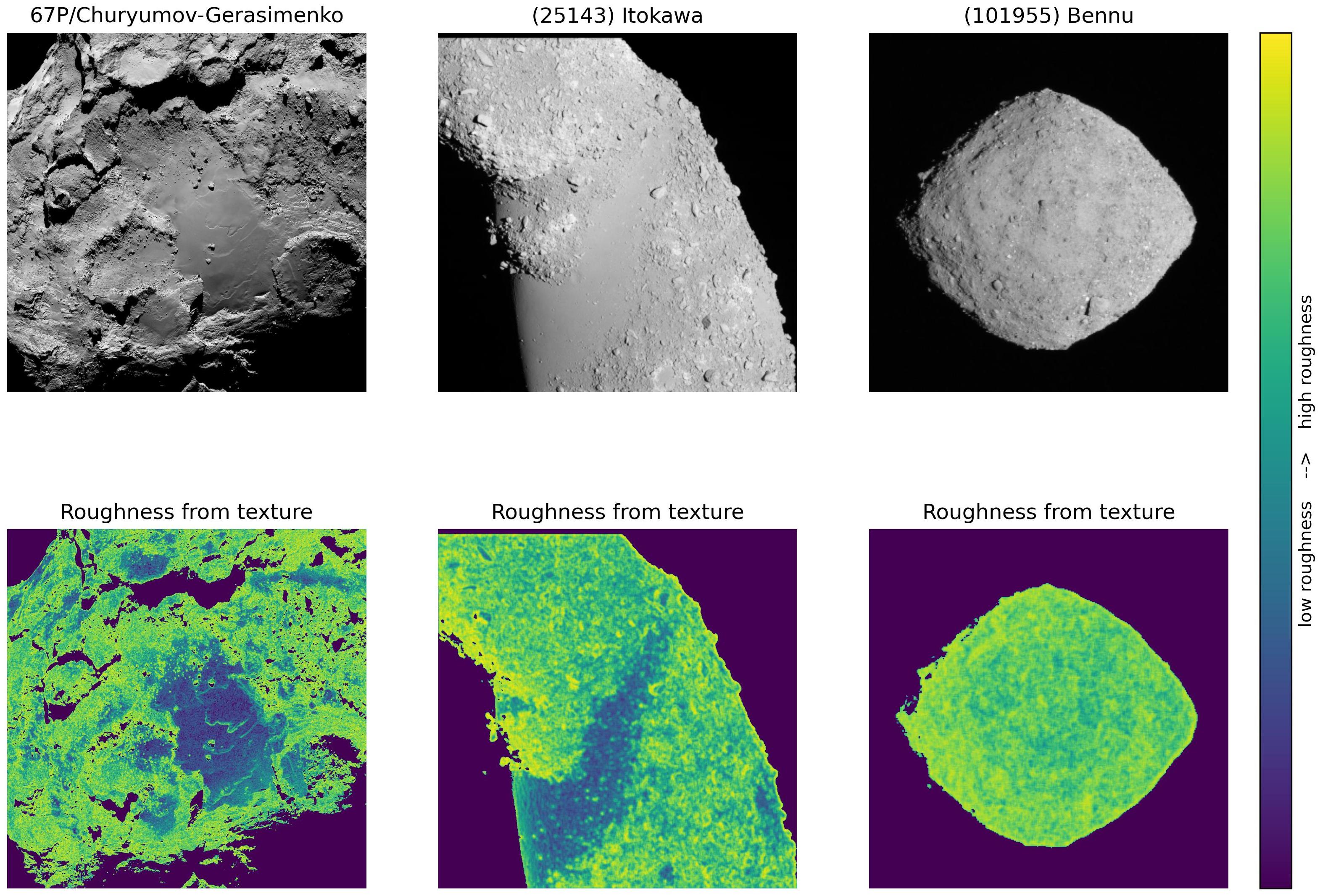}
\end{center}

\textbf{Question:} Which small body in the solar system exhibits the most uniform surface roughness according to the entropy of information measured in the image?

\textbf{Option:} 
\begin{enumerate}[label=(\Alph*)]
    \item 67P/Churyumov-Gerasimenko
    \item (25143) Itokawa
    \item (101955) Bennu
    \item None of the above
\end{enumerate}

\textbf{Answer:} \textcolor{ForestGreen}{D}
\vspace{0.5em}

\textbf{Ovis2-34B:} \textcolor{ForestGreen}{D}
\vspace{0.5em}

\textbf{ChatGPT-4o: } \textcolor{ForestGreen}{D}
\vspace{0.5em}

\textbf{Doubao-1.5-vision-pro: } \textcolor{ForestGreen}{D}
\vspace{0.5em}

\textbf{InternVL3-38B: } \textcolor{ForestGreen}{D}
\vspace{0.5em}

\textbf{Qwen2.5-VL-72B: } \textcolor{ForestGreen}{D}
\vspace{0.5em}

\textbf{LLaVA\_Onevision\_72B: } \textcolor{ForestGreen}{D}
\vspace{0.5em}

\textbf{Gemma3-12B: } \textcolor{ForestGreen}{D}

\end{examplebox}
\begin{center}
Figure B7: Case 7 of AstroMMBench in EP subdomain.
\end{center}

\begin{examplebox}[White]{\text{Correct responses: 14/25 models}}
\small
\begin{center}
    \includegraphics[width=\textwidth, height=3.5cm]{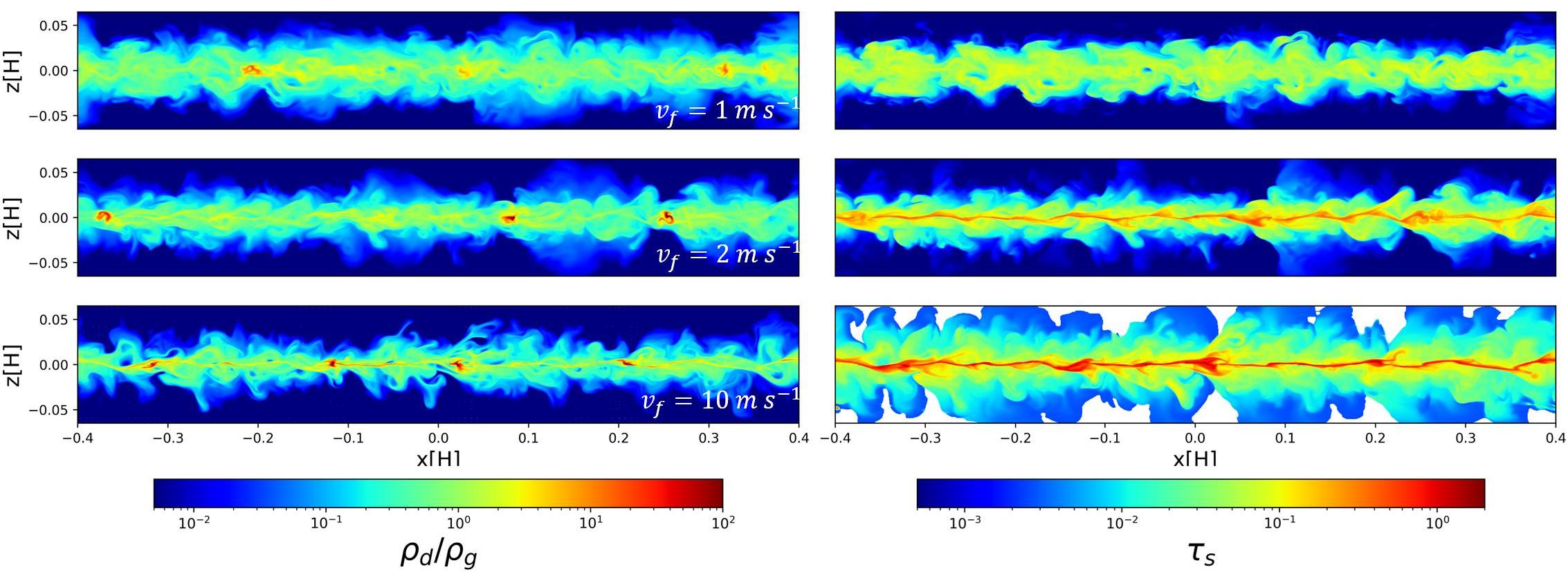}
\end{center}

\textbf{Question:} What is the primary effect of increasing the fragmentation velocity $v_f$ on the dust density distribution in the simulation?

\textbf{Option:} 
\begin{enumerate}[label=(\Alph*)]
    \item Formation of smaller dust clumps
    \item Decreased mass-averaged stopping time $\tau_s$
    \item Uniform distribution of dust
    \item Formation of larger dust clumps
\end{enumerate}

\textbf{Answer:} \textcolor{ForestGreen}{D}
\vspace{0.5em}

\textbf{Ovis2-34B:} \textcolor{ForestGreen}{D}
\vspace{0.5em}

\textbf{ChatGPT-4o: } \textcolor{ForestGreen}{D}
\vspace{0.5em}

\textbf{Doubao-1.5-vision-pro: } \textcolor{ForestGreen}{D}
\vspace{0.5em}

\textbf{InternVL3-38B: } \textcolor{red}{A}
\vspace{0.5em}

\textbf{Qwen2.5-VL-72B: } \textcolor{ForestGreen}{D}
\vspace{0.5em}

\textbf{LLaVA\_Onevision\_72B: } \textcolor{ForestGreen}{D}
\vspace{0.5em}

\textbf{Gemma3-12B: } \textcolor{red}{A}

\end{examplebox}
\begin{center}
Figure B8: Case 8 of AstroMMBench in EP subdomain.
\end{center}

\begin{examplebox}[White]{\text{Correct responses: 9/25 models}}
\small
\begin{center}
    \includegraphics[width=\textwidth, height=7cm]{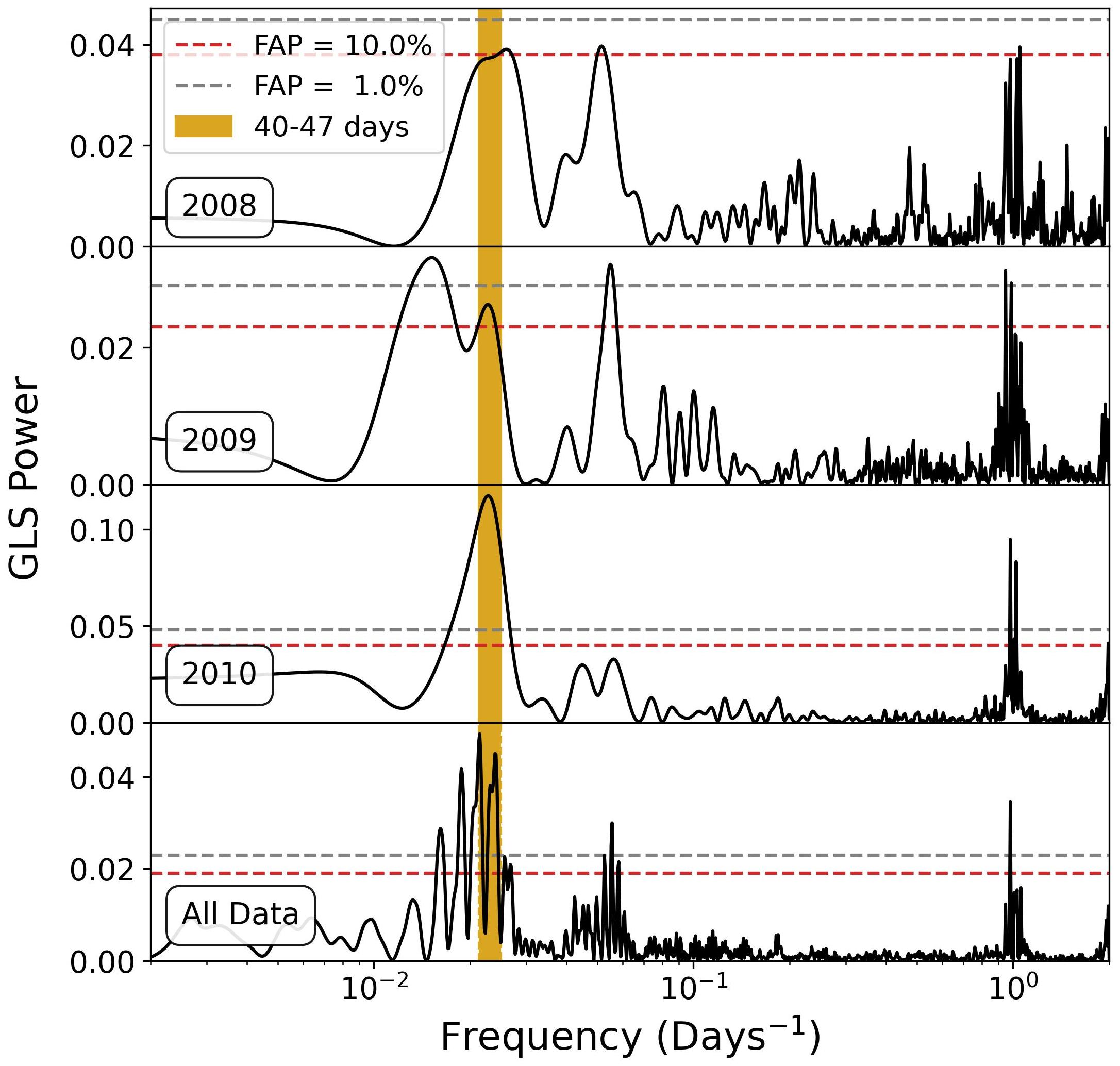}
\end{center}

\textbf{Question:} What is the most likely cause of the strong signal at approximately 40 days in the GLS periodogram of TOI-1450A?

\textbf{Option:} 
\begin{enumerate}[label=(\Alph*)]
    \item Planetary transit
    \item Stellar rotation
    \item Binary star system
    \item Instrumental artifact
\end{enumerate}

\textbf{Answer:} \textcolor{ForestGreen}{B}
\vspace{0.5em}

\textbf{Ovis2-34B:} \textcolor{red}{A}
\vspace{0.5em}

\textbf{ChatGPT-4o: } \textcolor{red}{A}
\vspace{0.5em}

\textbf{Doubao-1.5-vision-pro: } \textcolor{red}{A}
\vspace{0.5em}

\textbf{InternVL3-38B: } \textcolor{red}{A}
\vspace{0.5em}

\textbf{Qwen2.5-VL-72B: } \textcolor{ForestGreen}{B}
\vspace{0.5em}

\textbf{LLaVA\_Onevision\_72B: } \textcolor{red}{A}
\vspace{0.5em}

\textbf{Gemma3-12B: } \textcolor{ForestGreen}{B}

\end{examplebox}
\begin{center}
Figure B9: Case 9 of AstroMMBench in EP subdomain.
\end{center}

\subsection{Cosmology and Nongalactic Astrophysics (CO)}\label{appendixB:CO}

\begin{examplebox}[White]{\text{Correct responses: 22/25 models}}
\small
\begin{center}
    \includegraphics[width=\textwidth, height=12cm]{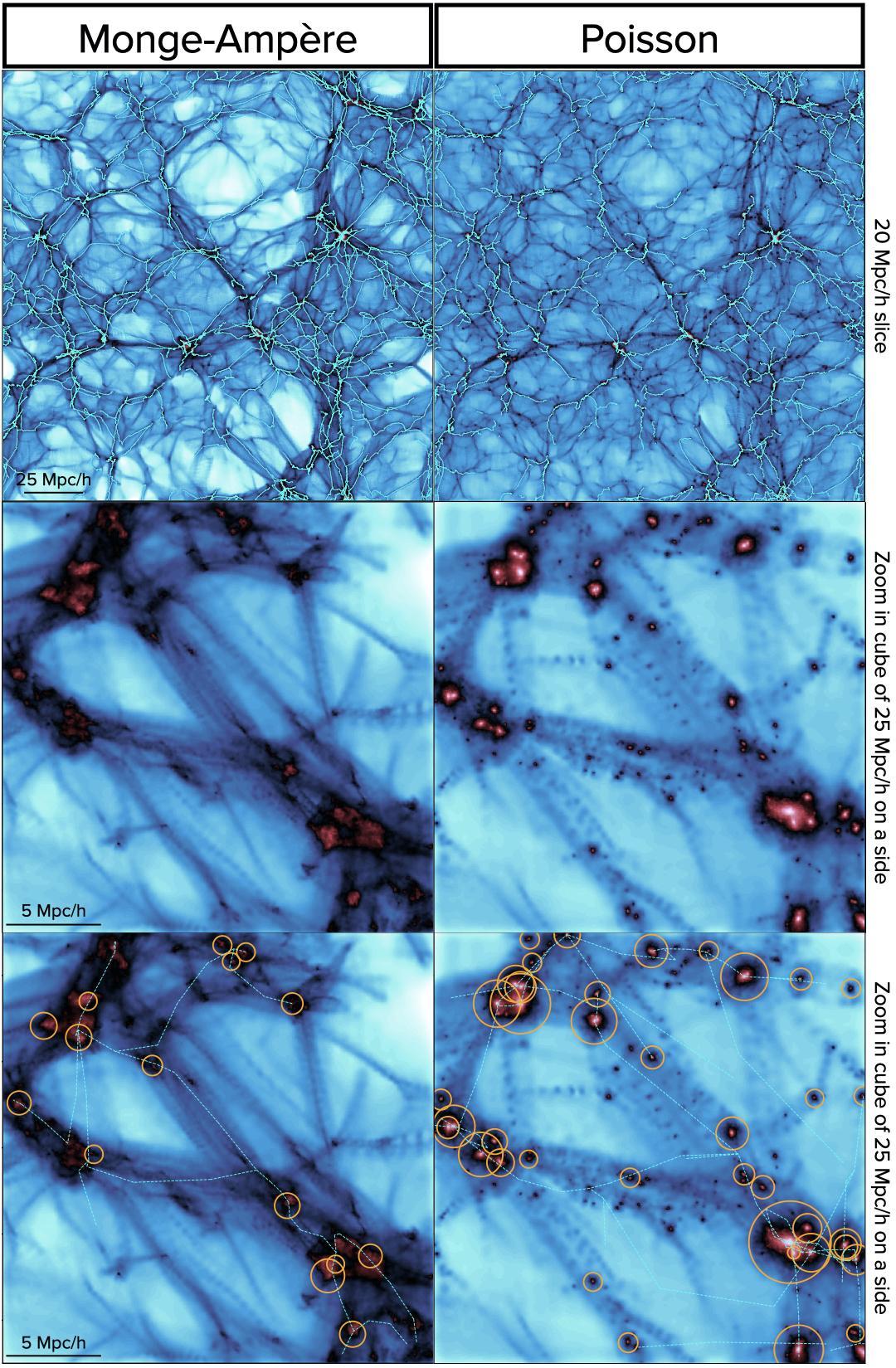}
\end{center}

\textbf{Question:} Which gravity theory, as depicted in the image, exhibits a more complex and abundant network of cosmic filaments?

\textbf{Option:} 
\begin{enumerate}[label=(\Alph*)]
    \item Poisson ($\Lambda$CDM)
    \item Monge-Ampère
    \item Both exhibit similar complexity
    \item Neither, the complexity is indistinguishable
\end{enumerate}

\textbf{Answer:} \textcolor{ForestGreen}{B}
\vspace{0.5em}

    \textbf{Ovis2-34B:} \textcolor{ForestGreen}{B}
\vspace{0.5em}

\textbf{ChatGPT-4o: } \textcolor{ForestGreen}{B}
\vspace{0.5em}

\textbf{Doubao-1.5-vision-pro: } \textcolor{ForestGreen}{B}
\vspace{0.5em}

\textbf{InternVL3-38B: } \textcolor{ForestGreen}{B}
\vspace{0.5em}

\textbf{Qwen2.5-VL-72B: } \textcolor{ForestGreen}{B}
\vspace{0.5em}

\textbf{LLaVA\_Onevision\_72B: } \textcolor{red}{A}
\vspace{0.5em}

\textbf{Gemma3-12B: } \textcolor{ForestGreen}{B}

\end{examplebox}
\begin{center}
Figure B10: Case 10 of AstroMMBench in CO subdomain.
\end{center}

\begin{examplebox}[White]{\text{Correct responses: 18/25 models}}
\small
\begin{center}
    \includegraphics[width=\textwidth, height=2.5cm]{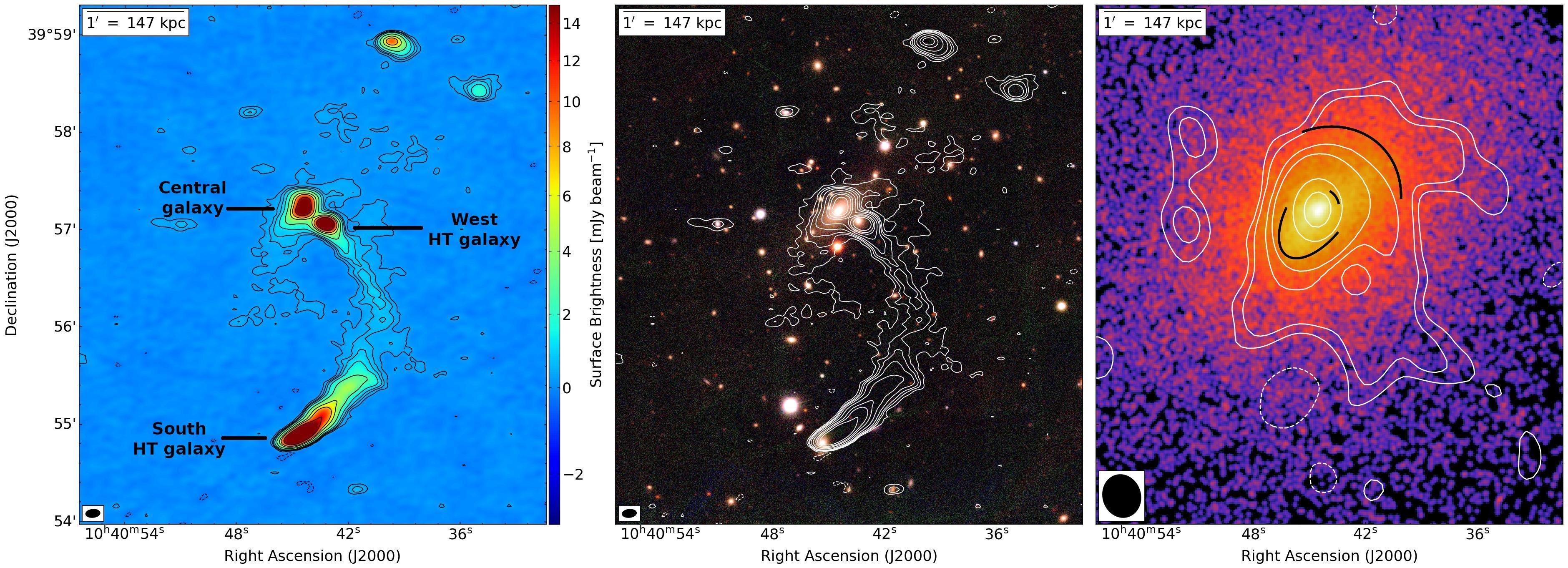}
\end{center}

\textbf{Question:} What is the primary feature indicated by the elongated X-ray emission in the right panel of the image?

\textbf{Option:} 
\begin{enumerate}[label=(\Alph*)]
    \item A single-armed spiral galaxy
    \item A region of high star formation activity
    \item A supermassive black hole
    \item A cold front in the galaxy cluster
\end{enumerate}

\textbf{Answer:} \textcolor{ForestGreen}{D}
\vspace{0.5em}

\textbf{Ovis2-34B:} \textcolor{ForestGreen}{D}
\vspace{0.5em}

\textbf{ChatGPT-4o: } \textcolor{ForestGreen}{D}
\vspace{0.5em}

\textbf{Doubao-1.5-vision-pro: } \textcolor{ForestGreen}{D}
\vspace{0.5em}

\textbf{InternVL3-38B: } \textcolor{ForestGreen}{D}
\vspace{0.5em}

\textbf{Qwen2.5-VL-72B: } \textcolor{ForestGreen}{D}
\vspace{0.5em}

\textbf{LLaVA\_Onevision\_72B: } \textcolor{ForestGreen}{D}
\vspace{0.5em}

\textbf{Gemma3-12B: } \textcolor{ForestGreen}{D}

\end{examplebox}
\begin{center}
Figure B11: Case 11 of AstroMMBench in CO subdomain.
\end{center}

\begin{examplebox}[White]{\text{Correct responses: 4/25 models}}
\small
\begin{center}
    \includegraphics[width=\textwidth, height=7.5cm]{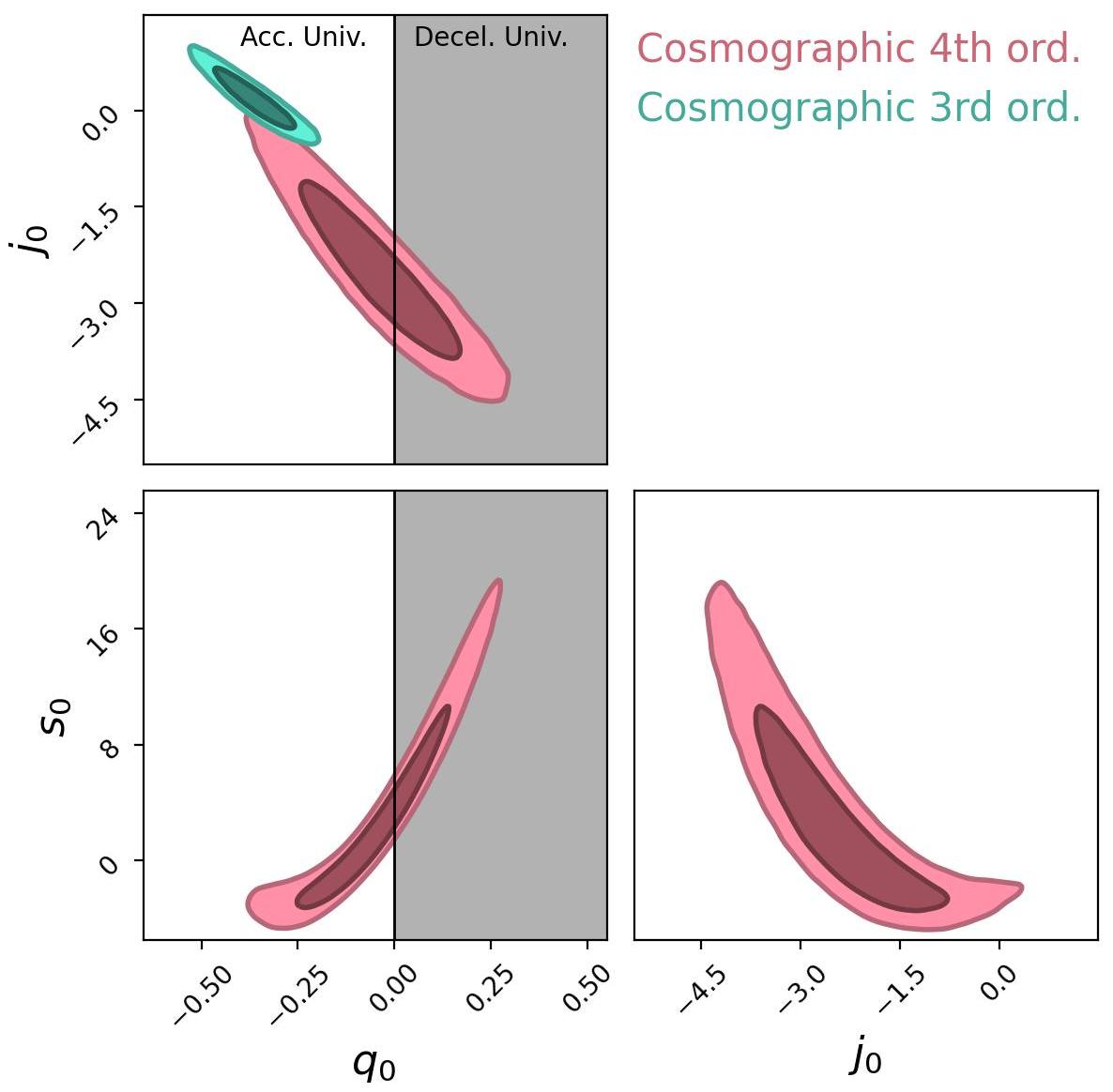}
\end{center}

\textbf{Question:} Which cosmographic model, as depicted in the image, provides stronger evidence for an accelerating universe?

\textbf{Option:} 
\begin{enumerate}[label=(\Alph*)]
    \item Cosmographic 3rd order model
    \item Cosmographic 4th order model
    \item Both models equally
    \item Neither model
\end{enumerate}

\textbf{Answer:} \textcolor{ForestGreen}{A}
\vspace{0.5em}

\textbf{Ovis2-34B:} \textcolor{ForestGreen}{A}
\vspace{0.5em}

\textbf{ChatGPT-4o: } \textcolor{ForestGreen}{A}
\vspace{0.5em}

\textbf{Doubao-1.5-vision-pro: } \textcolor{red}{B}
\vspace{0.5em}

\textbf{InternVL3-38B: } \textcolor{red}{B}
\vspace{0.5em}

\textbf{Qwen2.5-VL-72B: } \textcolor{red}{B}
\vspace{0.5em}

\textbf{LLaVA\_Onevision\_72B: } \textcolor{red}{B}
\vspace{0.5em}

\textbf{Gemma3-12B: } \textcolor{red}{B}

\end{examplebox}
\begin{center}
Figure B12: Case 12 of AstroMMBench in CO subdomain.
\end{center}

\subsection{Astrophysics of Galaxies (GA)}\label{appendixB:GA}

\begin{examplebox}[White]{\text{Correct responses: 23/25 models}}
\small
\begin{center}
    \includegraphics[width=\textwidth, height=2.5cm]{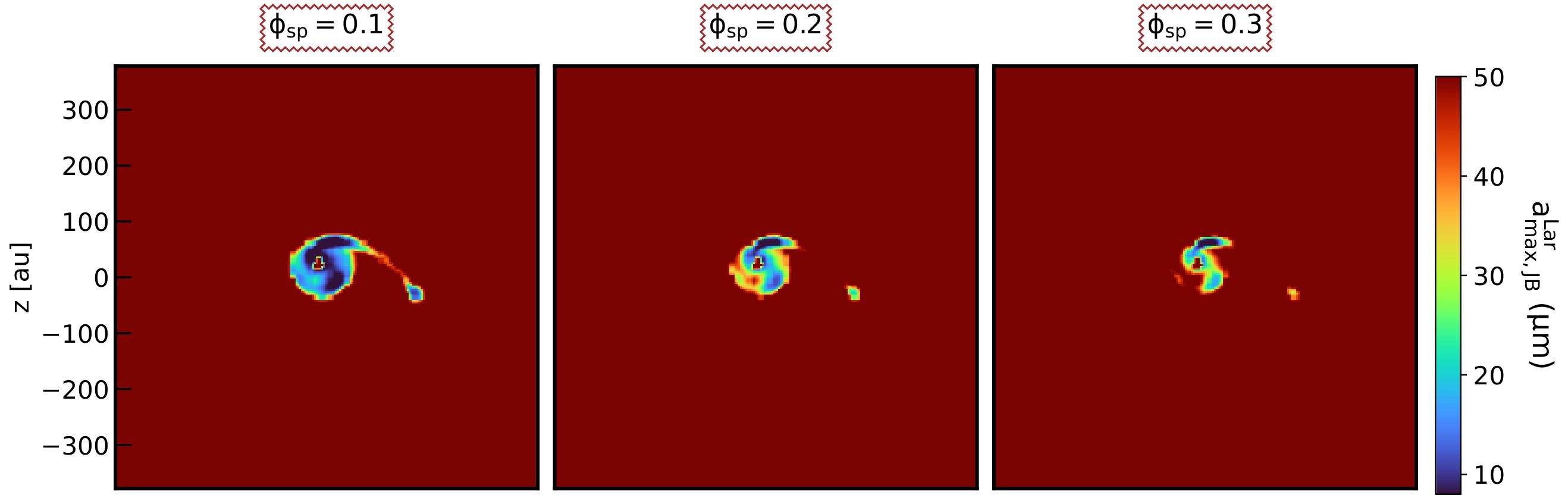}
\end{center}

\textbf{Question:} What is the primary effect of increasing the volume filling factor of iron clusters inside dust grains ($\phi_{\rm sp}$) on the distribution of dust grain sizes within 400 au of the disk midplane?

\textbf{Option:}
\begin{enumerate}[label=(\Alph*)]
    \item Decreased magnetic alignment of very large grains (VLGs)
    \item Increased internal alignment of dust grains
    \item Enhanced Magnetic Alignment by Radiative Torques (MRAT) alignment for micron-sized grains
    \item Reduced polarization degree within the disk scale
\end{enumerate}

\textbf{Answer:} \textcolor{ForestGreen}{C}
\vspace{0.5em}

\textbf{Ovis2-34B:} \textcolor{ForestGreen}{C}
\vspace{0.5em}

\textbf{ChatGPT-4o: } \textcolor{ForestGreen}{C}
\vspace{0.5em}

\textbf{Doubao-1.5-vision-pro: } \textcolor{ForestGreen}{C}
\vspace{0.5em}

\textbf{InternVL3-38B: } \textcolor{ForestGreen}{C}
\vspace{0.5em}

\textbf{Qwen2.5-VL-72B: } \textcolor{ForestGreen}{C}
\vspace{0.5em}

\textbf{LLaVA\_Onevision\_72B: } \textcolor{ForestGreen}{C}
\vspace{0.5em}

\textbf{Gemma3-12B: } \textcolor{red}{B}

\end{examplebox}
\begin{center}
Figure B13: Case 13 of AstroMMBench in GA subdomain.
\end{center}

\begin{examplebox}[White]{\text{Correct responses: 15/25 models}}
\small
\begin{center}
    \includegraphics[width=0.9\textwidth]{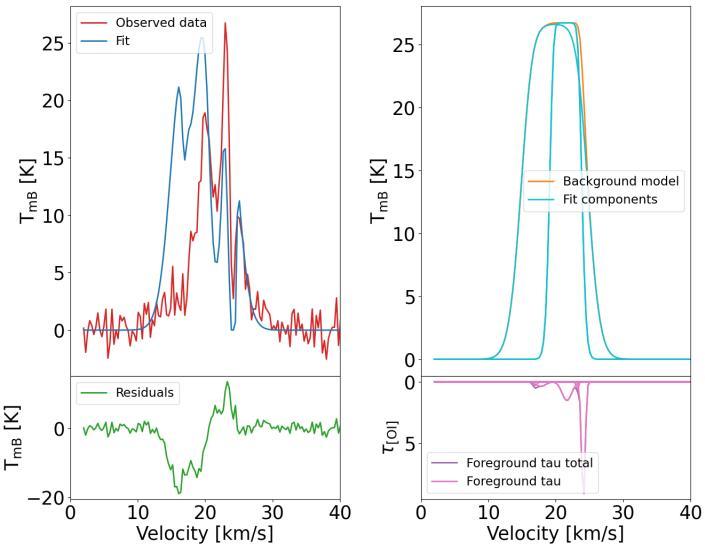}
\end{center}

\textbf{Question:} What is the primary reason for the significant residuals in the observed data fit shown in the image?

\textbf{Option:}
\begin{enumerate}[label=(\Alph*)]
    \item Insufficient data points
    \item Incorrect background model
    \item Fixed foreground parameters
    \item Instrumental error
\end{enumerate}

\textbf{Answer:} \textcolor{ForestGreen}{C}
\vspace{0.5em}

\textbf{Ovis2-34B:} \textcolor{ForestGreen}{C}
\vspace{0.5em}

\textbf{ChatGPT-4o: } \textcolor{ForestGreen}{C}
\vspace{0.5em}

\textbf{Doubao-1.5-vision-pro: } \textcolor{ForestGreen}{C}
\vspace{0.5em}

\textbf{InternVL3-38B: } \textcolor{ForestGreen}{C}
\vspace{0.5em}

\textbf{Qwen2.5-VL-72B: } \textcolor{ForestGreen}{C}
\vspace{0.5em}

\textbf{LLaVA\_Onevision\_72B: } \textcolor{ForestGreen}{C}
\vspace{0.5em}

\textbf{Gemma3-12B: } \textcolor{red}{B}

\end{examplebox}
\begin{center}
Figure B14: Case 14 of AstroMMBench in GA subdomain.
\end{center}

\begin{examplebox}[White]{\text{Correct responses: 1/25 models}}
\small
\begin{center}
    \includegraphics[width=0.85\textwidth]{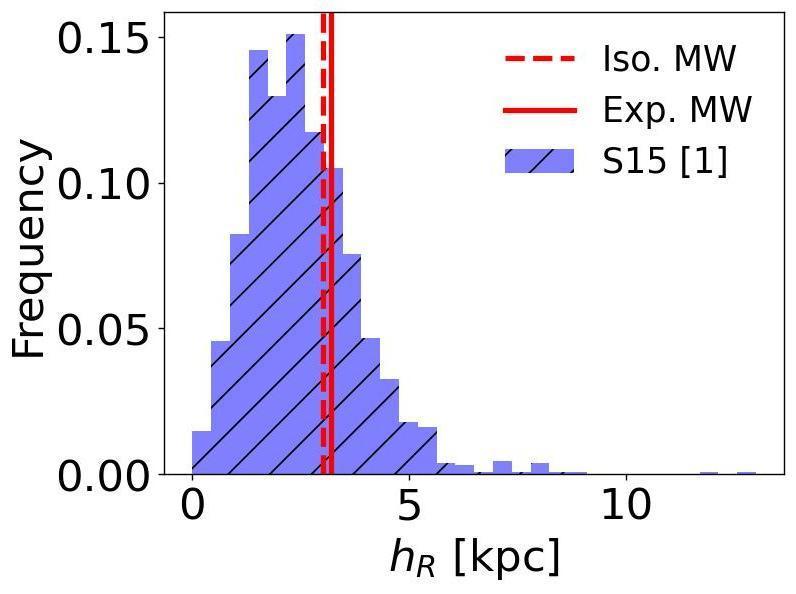}
\end{center}

\textbf{Question:} What is the primary purpose of the histogram in the image?

\textbf{Option:}
\begin{enumerate}[label=(\Alph*)]
    \item To compare the scale length of the Milky Way with other galaxies
    \item  To determine the frequency of galaxies with specific scale lengths
    \item To illustrate the distribution of galaxy types in the sample
    \item To show the relationship between scale length and galaxy mass
\end{enumerate}

\textbf{Answer:} \textcolor{ForestGreen}{A}
\vspace{0.5em}

\textbf{Ovis2-34B:} \textcolor{red}{B}
\vspace{0.5em}

\textbf{ChatGPT-4o: } \textcolor{red}{B}
\vspace{0.5em}

\textbf{Doubao-1.5-vision-pro: } \textcolor{red}{B}
\vspace{0.5em}

\textbf{InternVL3-38B: } \textcolor{red}{B}
\vspace{0.5em}

\textbf{Qwen2.5-VL-72B: } \textcolor{red}{B}
\vspace{0.5em}

\textbf{LLaVA\_Onevision\_72B: } \textcolor{red}{B}
\vspace{0.5em}

\textbf{Gemma3-12B: } \textcolor{ForestGreen}{A}

\end{examplebox}
\begin{center}
Figure B15: Case 15 of AstroMMBench in GA subdomain.
\end{center}

\subsection{High Energy Astrophysical Phenomena (HE)}\label{appendixB:HE}
\begin{examplebox}[White]{\text{Correct responses: 22/25 models}}
\small
\begin{center}
    \includegraphics[width=0.85\textwidth]{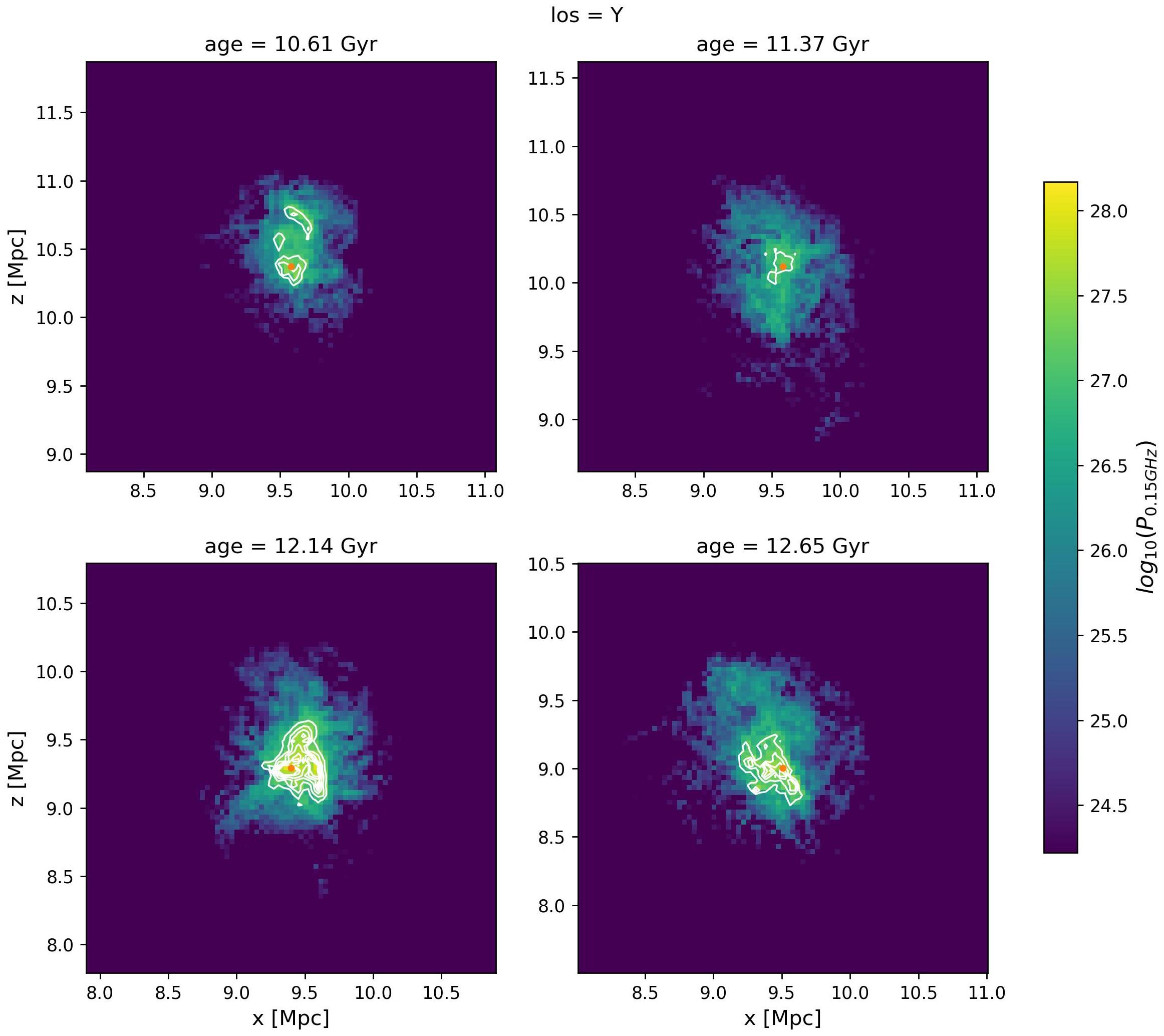}
\end{center}

\textbf{Question:} What is the primary reason for the asymmetric radio emission patterns observed in the maps?

\textbf{Option:} 
\begin{enumerate}[label=(\Alph*)]
    \item Galaxy rotation
    \item Galaxy cluster merger
    \item Stellar winds
    \item Black hole activity
\end{enumerate}

\textbf{Answer:} \textcolor{ForestGreen}{B}
\vspace{0.5em}

\textbf{Ovis2-34B:} \textcolor{ForestGreen}{B}
\vspace{0.5em}

\textbf{ChatGPT-4o: } \textcolor{ForestGreen}{B}
\vspace{0.5em}

\textbf{Doubao-1.5-vision-pro: } \textcolor{ForestGreen}{B}
\vspace{0.5em}

\textbf{InternVL3-38B: } \textcolor{ForestGreen}{B}
\vspace{0.5em}

\textbf{Qwen2.5-VL-72B: } \textcolor{ForestGreen}{B}
\vspace{0.5em}

\textbf{LLaVA\_Onevision\_72B: } \textcolor{ForestGreen}{B}
\vspace{0.5em}

\textbf{Gemma3-12B: } \textcolor{red}{D}

\end{examplebox}
\begin{center}
Figure B16: Case 16 of AstroMMBench in HE subdomain.
\end{center}

\begin{examplebox}[White]{\text{Correct responses: 9/25 models}}
\small
\begin{center}
    \includegraphics[width=\textwidth]{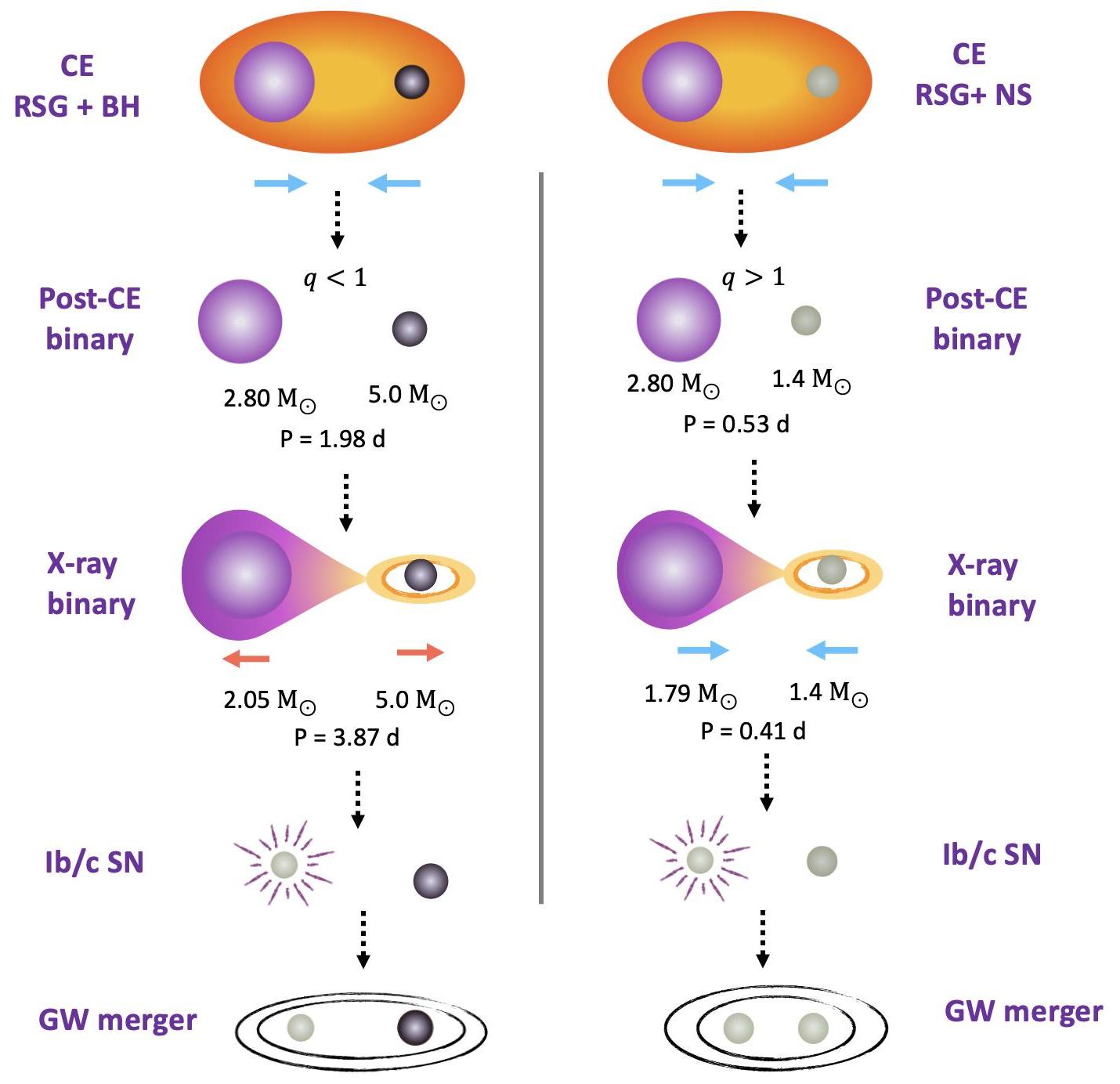}
\end{center}

\textbf{Question:} What is the final orbital period of the post-CE binary with a black hole (BH) companion after the mass transfer (MT) phase?

\textbf{Option:} 
\begin{enumerate}[label=(\Alph*)]
    \item 0.41 d
    \item 1.98 d
    \item 0.53 d
    \item 3.87 d
\end{enumerate}

\textbf{Answer:} \textcolor{ForestGreen}{D}
\vspace{0.5em}

\textbf{Ovis2-34B:} \textcolor{ForestGreen}{D}
\vspace{0.5em}

\textbf{ChatGPT-4o: } \textcolor{ForestGreen}{D}
\vspace{0.5em}

\textbf{Doubao-1.5-vision-pro: } \textcolor{ForestGreen}{D}
\vspace{0.5em}

\textbf{InternVL3-38B: } \textcolor{ForestGreen}{D}
\vspace{0.5em}

\textbf{Qwen2.5-VL-72B: } \textcolor{red}{B}
\vspace{0.5em}

\textbf{LLaVA\_Onevision\_72B: } \textcolor{red}{A}
\vspace{0.5em}

\textbf{Gemma3-12B: } \textcolor{red}{B}

\end{examplebox}
\begin{center}
Figure B17: Case 17 of AstroMMBench in HE subdomain.
\end{center}

\begin{examplebox}[White]{\text{Correct responses: 1/25 models}}
\small
\begin{center}
    \includegraphics[width=\textwidth]{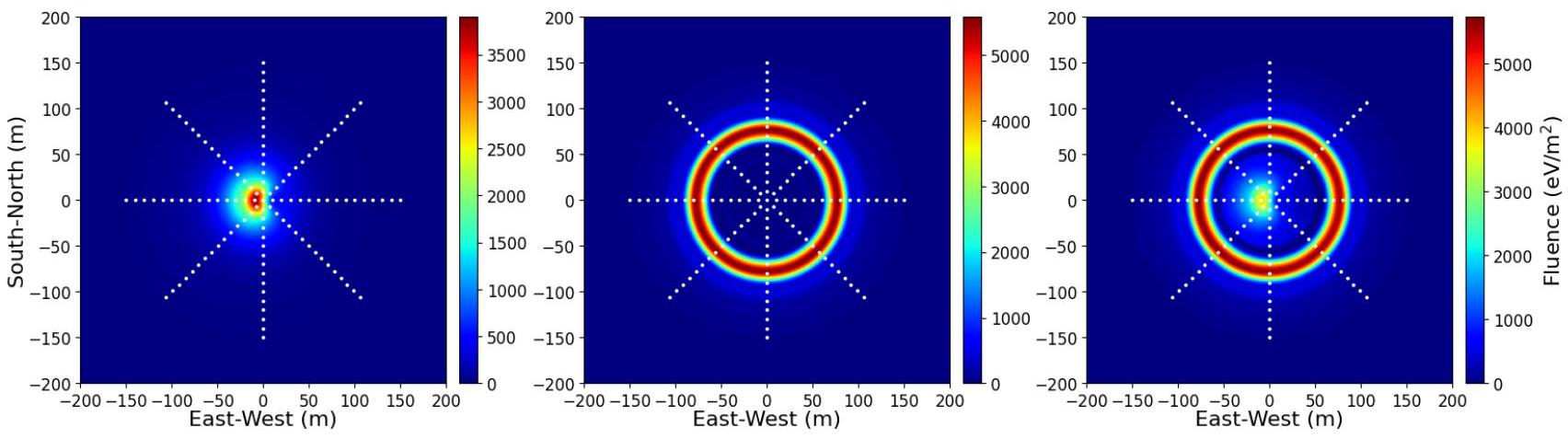}
\end{center}

\textbf{Question:} What is the primary reason for the distinct ring-like pattern observed in the middle panel of the image?

\textbf{Option:} 
\begin{enumerate}[label=(\Alph*)]
    \item Refraction of radio waves through the ice medium
    \item Reflection of radio waves off the ice surface
    \item Scattering of radio waves by ice crystals
    \item Interference between in-air and in-ice radio waves
\end{enumerate}

\textbf{Answer:} \textcolor{ForestGreen}{A}
\vspace{0.5em}

\textbf{Ovis2-34B:} \textcolor{red}{D}
\vspace{0.5em}

\textbf{ChatGPT-4o: } \textcolor{ForestGreen}{A}
\vspace{0.5em}

\textbf{Doubao-1.5-vision-pro: } \textcolor{red}{D}
\vspace{0.5em}

\textbf{InternVL3-38B: } \textcolor{red}{D}
\vspace{0.5em}

\textbf{Qwen2.5-VL-72B: } \textcolor{red}{D}
\vspace{0.5em}

\textbf{LLaVA\_Onevision\_72B: } \textcolor{red}{B}
\vspace{0.5em}

\textbf{Gemma3-12B: } \textcolor{red}{D}

\end{examplebox}
\begin{center}
Figure B18: Case 18 of AstroMMBench in HE subdomain.
\end{center}

\end{document}